\documentclass[sigconf, nonacm]{acmart}

\usepackage{color, soul}
\usepackage{xcolor}
\usepackage[frozencache,cachedir=./_minted]{minted}
\usepackage[export]{adjustbox}
\usepackage{subcaption}
\usepackage{multirow}
\usepackage{hhline}
\usepackage{tabularx}
\usepackage{wasysym}
\usepackage{tikz}
\usepackage{makecell}
\usepackage{enumitem}

\definecolor{alex-color}{rgb}{0.57, 0.910, 0.74}

\definecolor{stef-color}{rgb}{1, 0.898, 0.706}

\definecolor{ddkang-blue}{rgb}{0.69, 0.87, 0.94}

\definecolor{charith-color}{rgb}{1, 0.5, 0.5}

\colorlet{light-gray}{gray!10}
\definecolor{ForestGreen}{rgb}{0.13, 0.55, 0.13} 

\def\system{\textsc{PandasBench}}
\def\numnbs{102}
\def\numcells{3,721}
\def\numnbsfixed{114}
\def\numnbsthrownout{12}

\newcommand{\minihead}[1]{\textbf{\emph{#1}.}}
\newcommand{\code}[1]{\texttt{#1}}

\newcommand{\mycross}{\textcolor{red}{\textbf{$\times$}}}
\newcommand{\mycheck}{\textcolor{ForestGreen}{\textbf{\checkmark}}}

\newcommand{\mybullet}{

\raisebox{-0.3\height}{\begin{tikzpicture}[scale=0.27]
    \filldraw[orange] (0,0) -- (0.5,1) -- (1,0) -- cycle;
    \node[white, font=\bfseries, scale=0.8] at (0.5,0.4) {!}; 
\end{tikzpicture}}
}

\begin{document}
\title{\system{}: A Benchmark for the Pandas API}

\author{Alex Broihier}
\authornote{Both authors contributed equally to this research.}
\affiliation{%
  \institution{University of Illinois (UIUC), U.S.A}
}
\email{adb12@illinois.edu}

\author{Stefanos Baziotis}
 \authornotemark[1]
\affiliation{%
  \institution{University of Illinois (UIUC), U.S.A}
}
\email{sb54@illinois.edu}

\author{Daniel Kang}
\affiliation{%
  \institution{University of Illinois (UIUC), U.S.A}
}
\email{ddkang@illinois.edu}

\author{Charith Mendis}
\affiliation{%
  \institution{University of Illinois (UIUC), U.S.A}
}
\email{charithm@illinois.edu}

\begin{abstract}
The Pandas API has been central to the success of \code{pandas} and its
alternatives. Despite its importance, there is no benchmark for it, and we argue that we cannot repurpose existing benchmarks (from other domains) for the Pandas API.

In this paper, we introduce requirements that are necessary for a Pandas API
benchmark, and present the first benchmark that fulfills them: \system{}. We
argue that it should evaluate the \textit{real-world coverage} of a technique.
Yet, real-world coverage is not sufficient for a useful benchmark, and so we also: cleaned it from irrelevant code, adapted it for benchmark usage,
and introduced input scaling. We claim that uniform scaling used in other
benchmarks (e.g., TPC-H) is too coarse-grained for \system{}, and use a
non-uniform scaling scheme. \system{} is the largest Pandas API benchmark to
date, with \numnbs{} notebooks and \numcells{} cells.

We used \system{} to evaluate Modin, Dask, Koalas, and Dias. This is the
largest-scale evaluation of all these techniques to date. Prior works report significant speedups
using constrained benchmarks, but we show that on a larger benchmark with real-world code, the most notebooks that got a speedup were 8/102 (\textasciitilde8\%) for Modin, and 0 for both Koalas and Dask. Dias showed speedups in up to 55 notebooks (\textasciitilde54\%), but it rewrites code incorrectly in certain cases, which had not been observed in prior work. Second, we identified many failures: Modin runs only 72/102 (\textasciitilde70\%) notebooks, Dask 4 (\textasciitilde4\%), Koalas 10 (\textasciitilde10\%), and Dias 97 (\textasciitilde95\%).
\end{abstract}

\maketitle

\ifdefempty{\vldbavailabilityurl}{}{
\begingroup\small\noindent\raggedright\textbf{Artifact Availability:}\\
All the source code and data required to reproduce the experiments can be found
at \url{https://github.com/ADAPT-uiuc/PandasBench}.
\endgroup
}

\section{Introduction}

The sharp rise in the popularity of data science \cite{data_science_popular} has
been accompanied by a rapid growth of \code{pandas}, driven by users
ranging from business analysts \cite{pandas_in_business} to social scientists
\cite{pandas_in_social}. Prior work~\cite{analysis_millions_notebooks} that analyzed over a million notebooks found
that 42.3\% (of those that imported packages) imported \code{pandas}, topped by
no other data-science library. Some reasons for this growth include the
flexibility of \code{pandas}, the ability to work within a notebook environment,
and the interoperability with other libraries \cite{dias}.

These benefits are characteristics of the Pandas \textit{API}, of which
\code{pandas} is only an implementation. This implementation is secondary and
has so many limitations that have led to numerous Pandas API optimization
techniques \cite{dias,scirpy,modin_second,dask,koalas,magpie,ponder}.

\minihead{No Pandas API benchmark exists} Despite the significance of the Pandas
API and the advent of these techniques, \textit{no high-quality benchmark is
used in prior work}. Existing Pandas optimization techniques have been
evaluated using code that is too small, and/or doesn't include real-world code.
Weld \cite{weld} used only 6 Pandas notebooks to evaluate its improvement over
\code{pandas}. Modin---probably the most used Pandas alternative---used only
\textit{four, synthetic} benchmarks, each testing an \emph{individual}
operation~\cite{modin_second}. However, almost no real-world Pandas API use case
(e.g., notebooks found on Kaggle, Github, etc.) includes just a single
operation. Most Pandas API workloads are diverse, ad-hoc, and they include
interactions of the API with both other libraries and pure Python \cite{dias}. 

Experimental evidence suggests that this discrepancy is important in practice.
While Modin shows multi-fold speedups~\cite{modin_second} over \code{pandas} for
isolated operations (e.g., \code{fillna()}), when it is tested over whole
real-world notebooks, it slows down 19/20 of them.\footnote{Modin is intended
for scaling out, which is different than the setting explored by
Dias~\cite{dias}. But the experimental setups of the two evaluations we cite are
almost the same. In the Modin paper~\cite{modin_second} Modin shows significant
speedups in 8 and 16 cores in \textit{isolated operations}, whereas in the Dias
paper~\cite{dias} it shows significant slowdowns 8 and 12 cores in
\textit{real-world snippets}. The machine used in Dias was different than
Modin's; that is, the Modin machine had 128 cores and 3TB of memory, while the
Dias machine had 12 cores and 32 GB of memory. But the difference is relevant
only for higher core counts (e.g., Modin also shows speedups for 128 cores) and
larger dataset sizes. Dias uses datasets of about 2GB. So, in this case the only
variable is the code used.} In addition, Modin is not an isolated case. Koalas
slows down real-world \code{Pandas} snippets up to $155\times$ \cite{dias} and
in Appendix~\ref{app:non-pandas} we show that while Dask \cite{dask} performs
well on TPC-H queries, it slows down commonly occurring Pandas snippets. Even
Dias \cite{dias}, which has the largest Pandas API evaluation, uses only 20
notebooks. Our evaluation on a larger benchmark (\S\ref{sec:eval}) shows that
Dias rewrites code incorrectly and gives lower speedups. In short, such small
numbers are simply not enough to avoid a biased evaluation.

The reason that prior work did not use a high-quality Pandas benchmark is that
\textit{no such benchmark exists!} We gather relevant collections of Pandas code
in Table~\ref{tab:relwork_pandas}. While there are prior works that gather large
amounts of notebooks
\cite{kgotrrent,distill_kaggle,chi_dataset,data_science_looking_glass,datalore_dataset},
unfortunately, most are not executable. The few that are (the code used in
Dias~\cite{dias} and Modin~\cite{modin_second}) are too small. 

\textbf{\emph{Can we repurpose existing benchmarks?}} We could attempt to
repurpose existing benchmarks, but these are insufficient too for the Pandas
API. Dataframes are closest to matrices and the relational model~\cite{modin_first}, but their
benchmarks are insufficient. For example, dataframes use both strings
and numbers, while matrices only numbers. This is also why \code{numpy}
benchmarks \cite{npbench,numpy_benches, pythran_tests,numba_bench} do not
suffice; because they target numerical processing.

Dataframes are also different from relational tables. For example,
TOLABELS~\cite{modin_first} (i.e., setting a column as an index of the
dataframe, which in the Pandas API happens usually with \code{set\_index()}),
followed by a transpose ``is impossible using relational
operators''~\cite{modin_first}. Yet, we explore the TPC
family\cite{tpc_summaries} further. First, no variant uses the Pandas API.
Further, none includes any of the following characteristics, which are common in
Pandas API workloads: burst computations, reading and writing (raw data) files,
data cleaning, inspection (or visualization) of intermediate results, and
interaction with a host language. This is true for ProcBench \cite{procbench}
too, even though it includes procedural code and user-defined functions. In
Appendix~\ref{app:non-pandas} we give a more detailed argument on TPC-H.
Arguments for the other variants follow similarly.

Thus, we believe that a standard benchmark for the Pandas API is needed. In this
work, we focus on \textit{single-machine} workloads. This is because
millions of notebooks, including hundreds of thousands of notebooks on Github
\cite{analysis_millions_notebooks}, Kaggle \cite{Kaggle, dias}, Google Colab
\cite{colab_eda}, and all the workloads that are carried out on personal
laptops, use \code{pandas} on a \textit{single machine}. Two reasons explain
such workloads. First, \code{pandas} is single-threaded \cite{modin_first}, so
using multiple machines (or even multiple cores) is pointless. Second, managing
distributed clusters is challenging and time-consuming for Pandas API users~\cite{dias}. We now focus on two important characteristics that we argue any Pandas API benchmark should have.

\minihead{Code is not enough} As we need to be able to \emph{execute} code, we
also need valid input data, which we can't synthesize (see \S\ref{sec:collect}).
This is why the large collections in Table~\ref{tab:relwork_pandas} cannot be
extended.

\minihead{Real-world coverage} A Pandas optimization technique achieves
higher coverage the larger part of the API it supports. In this work, we highlight \emph{real-world coverage}, which
is a technique's coverage in practice, because e.g., in practice only 10\% of an
API may be used. For example, it's much less of an issue if a technique does not
support an obscure, rarely used operation (e.g., \code{.pop()}),
compared to not supporting e.g., \code{.sort\_values()}, which is found in
approximately 50\% of the notebooks. Moreover, there is a trade-off between
coverage and performance. In particular, the lower coverage, the smaller the subset, the easier it is to optimize it.

Real-world coverage is an issue in the available Pandas API alternatives. Prior work \cite{dias} sampled 20 random
notebooks from Kaggle, and found that six (30\%) were unable
to run with Modin. Dask and Koalas faced problems even with small snippets.

\begin{table}
  \centering
  \normalsize
  \caption{Characteristics of Pandas API collections. \system{} is the first
benchmark suite that marks all of them.}\label{tab:relwork_pandas}
  \begin{tabular}{c|c|c|c}
\toprule
\makecell{Collection} & \makecell{Real-World\\Code} &
\makecell{Executable/\\Available} & \makecell{Large \& \\ Diverse}\\
\midrule
DistillKaggle \cite{distill_kaggle}                      & \mycheck & \mycross & \mycheck \\
Datalore \cite{datalore_dataset}                         & \mycheck & \mycross & \mycheck \\
Looking Glass \cite{data_science_looking_glass}          & \mycheck & \mycross & \mycheck \\
KGTorrent \cite{kgtorrent}                               & \mycheck & \mycross & \mycheck \\
\makecell{Millions \cite{analysis_millions_notebooks}}   & \mycheck & \mycross & \mycheck \\
\texttt{modin} \cite{modin_second}                       & \mycross & \mycross & \mycross \\
Dias \cite{dias}                                         & \mycheck & \mycheck & \mycross \\
\textbf{\system{}}                                       & \mycheck & \mycheck & \mycheck \\
\bottomrule
  \end{tabular}
\end{table}

Furthermore, we argue that real-world coverage, while \textit{necessary}, is
\textit{not sufficient} to have a useful Pandas API benchmark. For instance,
many notebooks include a lot of other code that is not relevant for a Pandas API
benchmark. Machine-learning and plotting code are common examples. Thus, we need
to \textit{clean} them from such code. Also, some notebooks are not suitable for
a benchmark suite e.g., because they cannot be executed repeatedly with the same
behavior. So, we need to \textit{adapt} them so that they are appropriate for a
benchmark suite. Finally, most notebooks don't run long enough to get any
meaningful measurements. Others run for too long for e.g., cases where the
runtime does not matter (e.g., when gathering statistics of which API methods
are called). So, to cover both use cases, we need to provide an infrastructure
to \textit{scale} the data; both up and down. As we explain later
(\S\ref{sec:prep-scale}), this is much harder than conventional benchmarks.

In summary, we need a benchmark suite for Pandas API optimization techniques
that:

\begin{itemize}
  \item[\textbf{RWC}] Evaluates their \emph{real-world coverage}.
  \item[\textbf{REL}] Includes only relevant code.
  \item[\textbf{APPR}] Is appropriate for benchmark usages.
  \item[\textbf{DSC}] Allows effective data scaling.
\end{itemize}

As Table~\ref{tab:relwork_pandas} shows, existing Pandas API collection don't
even fulfill the \textbf{RWC} requirement, let alone all four. As we explain in
\S\ref{sec:collect} and \S\ref{sec:prep}, there are many challenges in
satisfying these requirements, which probably explains the current landscape.

We introduce, to the best of our knowledge, the first systematic effort to
create a benchmark that fulfills all 4 requirements: \system{}. We describe how
we collected notebooks from Kaggle (\S\ref{sec:collect}), fixed them
(\S\ref{sec:prep-fix}), cleaned them (\S\ref{sec:prep-clean}) and adapted them
(\S\ref{sec:prep-adapt}), and we introduce a practical non-uniform data scaling
scheme (\S\ref{sec:prep-scale}). After exploring \system{}
(\S\ref{sec:explore}), we used it to evaluate \code{pandas}, Modin
\cite{modin_second}, Dask \cite{dask}, Koalas \cite{koalas}, and Dias
\cite{dias} (\S\ref{sec:eval}). We summarize our contributions below:
\begin{itemize}[leftmargin=*]
  \item We introduce the largest executable benchmark suite for the Pandas API,
  composed of real-world, randomly picked notebooks.
  \item We fix, clean, and adapt these notebooks for benchmark use, we provide
  their input data and we allow non-uniform scaling.
  \item We perform the first systematic evaluation of real-world coverage, along
  with an evaluation of performance, of Modin~\cite{modin_second},
  Dask~\cite{dask}, Koalas~\cite{koalas}, Dias~\cite{dias}.
\end{itemize}

We believe that such a benchmark will allow people to evaluate more Pandas techniques across multiple dimensions (e.g., coverage,
performance), and that it will uncover improvement opportunities.
\section{Background \& Overview}

We first define terms that we use throughout the paper, and then a short overview of the rest of the paper.

\subsection{Definitions}
\label{sec:back-defs}

\textbf{\emph{On the word ``pandas.''}} Throughout the paper, we use the word
``pandas'' in lowercase and in typewriter font---i.e., \code{pandas}---to refer
to the popular \textit{library}, and its particular implementation. This is
generally used as a noun. We use the word ``pandas'' in uppercase and in normal
font---i.e., Pandas---to refer to anything related to the Pandas ecosystem,
e.g., the Pandas API, which is not necessarily directly related to the
\code{pandas} implementation. This is generally used as an adjective. We don't
need a similar distinction for other libraries, so we treat them as proper names
(e.g., Modin).

\textbf{\emph{Pandas API Techniques}.} We define a \emph{Pandas API alternative}, or simply \emph{Pandas alternative},
to be any implementation of the Pandas API other than the reference
implementation, which is that of \code{pandas}, while preserving the semantics
of the reference implementation. For example, Modin \cite{modin_first} is a
Pandas alternative while PolaRS is not because it implements almost none of the
Pandas API. In more detail, the set of Pandas alternatives we consider in this
paper contains: Modin \cite{modin_second}, Dask \cite{dask}, and Koalas
\cite{koalas}. All these alternatives aim to provide an optimized
implementation.

There are also other techniques that aim to optimize the Pandas API, not by
providing an alternative implementation, but by \emph{complementing} existing
implementations. We call these \emph{Pandas API complements}. In this work, we
consider Dias \cite{dias} and SCIRPy \cite{scirpy}.

We refer to the union of Pandas alternatives and Pandas complements as
\emph{Pandas API optimization techniques}, or simply \emph{techniques.}

\textbf{\emph{Collections}.} We use the generic term \emph{collection} to refer
to any set of code snippets of any size. We reserve the term \emph{benchmark}
for the collections that are \emph{executable}. We say that a collection is
\emph{diverse} if its elements were chosen randomly from real-world code. We dub
the term \emph{real-world coverage} and we say that a collection evaluates
real-world coverage if it:

\begin{enumerate}[label=\textbf{RWC\arabic*}]
  \item Uses real-world code, and
  \item Is executable, and
  \item Is large and diverse
\end{enumerate}

We now argue why the three conditions outlined above are necessary and
sufficient for real-world coverage. We need real-world code because that is what
is written in practice, but it is not enough. The collection should also be
executable, because in practice code runs. It should also be large and diverse.
Having only one of these two is not enough. For example, we may have three code
snippets that are very different from each other, and thus this collection is
diverse. But three code snippets is not enough to cover a large percentage of
real-world cases. Similarly, if a collection is large but has similar elements,
then it is the same as being small.

Finally, we may refer to a technique's real-world coverage. This is a different
(but only slightly) usage of the term, which is the coverage of the technique
(i.e., how much code it can run) over a benchmark that evaluates real-world
coverage.

\subsection{Overview}

\begin{figure}[ht]
  \centering
  \includegraphics[width=\columnwidth]{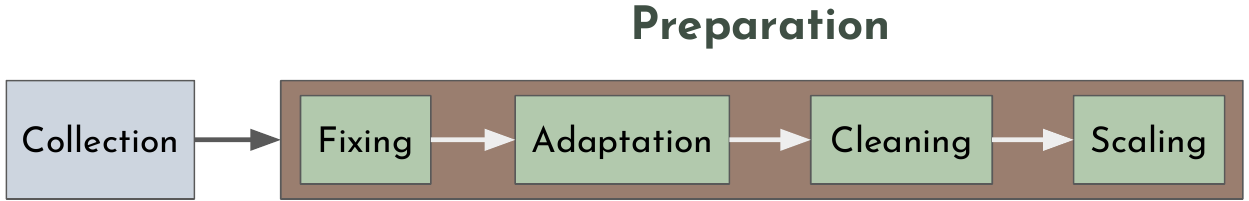}
  \caption{An overview of the construction of \system{}.}
  \label{plt:overview}
\end{figure}

In the remainder of the paper, we describe how \system{} was formed and we
present one use case: evaluating different Pandas optimization
techniques. In \S3 we describe how we collected the
notebooks. In \S4 we describe how we prepared the notebooks to fix them, clean them, adapt them, and scale their data. Figure~\ref{plt:overview} visualizes \S3 and \S4. In \S5 we explore \system{} to uncover interesting
characteristics. In \S6 we detail how we used \system{} to evaluate
different Pandas API techniques. We finish with \S7, in which we take a
closer look at related work.
\section{Notebook Collection}
\label{sec:collect}

In this section we describe how and from where we collected notebooks, so that we can achieve real-world coverage.

The hardest condition of real-world coverage (\S\ref{sec:back-defs}) to satisfy
is \textbf{RWC2}, i.e., to be able to execute the notebooks. To do that, we need
valid input data that the notebook can run on.\footnote{This data need not be
the original input data that the notebook used, but it needs to be valid. That
is, the notebook execution should not fail because of the input data.} We don't
know of any solution to the problem of \emph{generating} valid data for Pandas
API code. The closest solution we know is Pandera~\cite{pandera}, which solved
this issue in version \code{0.6.0}~\cite{pandera_issue}, but their solution
\textit{requires a \code{pandera} schema}. This is much more demanding than a
conventional schema (which assigns types to columns), and in the Pandas API we
don't even know the latter (see~\cite{modin_first} and \S\ref{sec:expl:input}).
Thus, our source needs to provide the original data that the notebook was
written for (apart from the code itself). A source that provides both notebooks
and their data is Kaggle~\cite{Kaggle}. Specifically, Kaggle provides
\textit{single-machine workloads}, which we target in this work.

Kaggle helps us satisfy condition \textbf{RWC1} because it hosts real
competitions (e.g., money is at stake) and popular datasets. So, the code that
people write to win these competitions, or to manipulate these datasets, is the
closest we can get to real-world code.

Finally, to satisfy condition \textbf{RWC3}, we downloaded \numnbs{} notebooks
\textit{randomly}, which amount to a total of \numcells{} code cells. We
collected notebooks satisfying the following criteria. First, it had to be a
Jupyter notebook. Kaggle hosts other kinds of workloads (e.g., Python (i.e.,
\code{.py}) or R files), but in this work, we only focus on notebooks. Second,
the code had to be in Python. Third, the code had to make heavy use of the
Pandas API. We believe that if 20\% of a notebook's \code{static} calls are from
the Pandas API, it makes a sufficiently heavy use of Pandas (especially
considering that Pandas is almost never used alone). This, however, does not
mean that the notebook needs to use \code{pandas}; theoretically, it can also
use e.g., Modin. But all the notebooks we selected use \code{pandas}. The
notebook also needs to have at least 10 cells, and we excluded all (20)
notebooks from Dias~\cite{dias} to avoid any bias in the evaluation.

Finally, as we mentioned, a notebook in \system{} needs to be executable.
However, we don't test/discard \emph{yet} whether we can execute the notebook.
This is because the notebook may contain other irrelevant code because of which
it may fail to run without modification. So, we defer checking whether the
notebook runs to notebook preparation.
\section{Notebook Preparation}
\label{sec:prep}

Most notebooks in our collection are not executable as they get downloaded from
Kaggle (e.g., the code does not match the input.). Thus, to satisfy condition
\textbf{RWC2} (\S\ref{sec:back-defs}), we need to \textit{fix} the
notebooks (and we remind the reader that the other two conditions are satisfied
already based on the way we collected notebooks).

However, as we argued in the introduction, real-world coverage is
\textit{necessary} but \textit{not sufficient} to have a useful Pandas API
benchmark. We also need to clean them---in the second stage, to achieve
\textbf{REL}---, adapt them---in the third stage, to achieve \textbf{APPR}.
Finally, to achieve \textbf{SCALE}, we explain in detail our non-conventional
decisions for data scaling.

In the subsections \S\ref{sec:prep-fix}-\S\ref{sec:prep-scale}, we explain in
more detail the challenges that necessitate each stage, and then
(\S\ref{algorithm}) we describe how we handled them. In total, we fixed and
cleaned \numnbsfixed{} notebooks and discarded \numnbsthrownout{} notebooks
after applying the process in \S\ref{algorithm}.

\subsection{Fixing}
\label{sec:prep-fix}


\minihead{Deprecated Methods} Kaggle doesn't provide information about the
original environment in which a notebook was created, e.g., library versions.
Thus, we need to guess these versions. We first attempt to execute the notebook
with \code{pandas-1.5.1}.\footnote{Most Kaggle notebooks have been written using
a \code{1.*} version of \code{pandas}. Version 1.5.1 was used in prior
work~\cite{dias} and is relatively new in \code{1.*}, before the switch to
\code{2.*}} A common problem arises when the code uses a deprecated method. Here is an example:

\begin{minted}[bgcolor=light-gray]{python}
# Original
X = trainData[testFeatures].fillna(0).as_matrix()
# Replacement
X = trainData[testFeatures].fillna(0).to_numpy()
\end{minted}

The \code{as\_matrix()} method is not available in our \code{pandas} version,
leading to an error. Thus, we need to rewrite it so that it uses the equivalent
\code{to\_numpy()} method.

\minihead{Input-Code Mismatch} Sometimes problems arise because the code does
not match the input data. One category is \textbf{Column Name Mismatch}, which
occurs when the input CSV has a column named differently than what the code
assumes. One reason such discrepancies happen is because the datasets used by
Kaggle notebooks have versions and the columns may have been slightly different
in an earlier version,\footnote{For example, you will see many versions of the
popular Titanic dataset, some with the columns in all lowercase, and some with
the first letter uppercase.} but we don't know that because we can't necessarily
know which dataset version the notebook used. Another case of mismatch we
encountered is \textbf{Column Size Mismatch}, in which we have code that looks
like \code{a['A'] = b}, where \code{a} is a dataframe and \code{b} is a
list-like object. Sometimes, the size of \code{b} does not match the size
expected by \code{a['A']}.

\subsection{Cleaning (REL)}
\label{sec:prep-clean}


\minihead{Dependencies} In general, cleaning is ``just'' deleting irrelevant
code. But other (relevant) code may depend on it. We show a real-world
example~\cite{NB_sudheer259_house-price-prediction} below, where some pandas
code is used to process values that results from machine-learning code:

\begin{minted}[bgcolor=light-gray]{python}
lm=smf.ols('...',data = train).fit()
lm.summary()
# This is relevant code because it's pandas code,
# but it depends on non-pandas code.
imc = pd.DataFrame(lm.pvalues)
\end{minted}

The problem of \textbf{Dependencies} was the core issue we faced during
cleaning. There are other problems we needed to deal with, like: there were
functions that used as parameters machine-learning values (and thus these parameters needed to be
removed); some return statements depended on non-pandas code; sometimes the
object over which a loop iterates depends (even if only partially) on non-pandas
code. But these problems, and in general \textit{all} the problems we faced
during cleaning, can be traced back to \textbf{Dependencies}. 

\subsection{Adaptation (APPR)}
\label{sec:prep-adapt}

\minihead{Repeated Execution} It is useful to be able to run a notebook multiple
times with the same behavior (e.g., for usability). However, this is not the
case for some notebooks out of the box. For example, some notebooks write to
disk to cache results. Thus, the notebook runs differently on the first run than
on subsequent runs.


\minihead{Lazy Evaluation}
Because we throw away plotting calls, for
libraries like Modin, Dask, etc. that use lazy evaluation, there are parts of the code that are not evaluated at all. Thus, we need to enforce it. Even though this is not necessary for \code{pandas}---and in notebook preparation we are concerned only with \code{pandas}---we do think it's a necessary property for a benchmark to make sure it evaluates what would get evaluated in the original notebook.

\subsection{Data Scaling (DSC)}
\label{sec:prep-scale}


We want to provide a capability for users to upscale and downscale the data. We now present the main challenges in that.

\minihead{Duplication is Naive} Data scaling is harder than just duplicating (or
cutting down) the already existing data (e.g., using \code{pd.concat()}). For
example, some notebooks need some columns not to have duplicates (e.g., because
they're used as an index); the code
below~\cite{NB_guyalmog-hr-analytics-predicting-employees-attrition} throws an
error if there are duplicate entries in the index columm:
\begin{minted}[bgcolor=light-gray]{python}
data[col2] == data.loc[i][col2]
\end{minted}

Other notebooks have code that depends on specific entries, which may not appear
if we naively downscale the data. As an example, this code~\cite{NB_charulatashelar/indian-startup-funding-exploratory-data-analysis} fixes
input data at a hardcoded index:

\begin{minted}[bgcolor=light-gray]{python}
data.loc[3029, 'Date'] = '22/01/2015'
\end{minted}

In addition, when upscaling the same input data
we need to make sure that we fix any duplicated instances of the same faulty
input data. Even the data type of a column may change with less data (as less
context is available for type inference). As such, we needed to go through each
notebook individually introduce special code to scale the data, satisfying any
necessary constraints. We scale the data \textit{once} before we start taking
any measurements.

\minihead{Non-Uniform Factor} Moreover, unlike most prior work, we don't use a
uniform scaling factor. For example, TPC-H uses the same scaling factor for all
data. However, unlike TPC-H, in \system{} every notebook uses different data and
in possibly different ways. More importantly, there is a large variance
of 1946.256 seconds in the running time of different notebooks. As such, instead of
targeting the scaling factor (e.g., providing infrastructure to scale all data
by $10\times$), we target running times. So, we implemented the infrastructure
such that we can provide individual scale factors.

\subsection{Preparation Process}
\label{algorithm}

In this section we present the process we followed to solve the problems presented in the previous sections. We named them in bold in these sections so that we can refer to them here. We applied the process manually since, as we hope will be clear below, we could not devise an algorithm that achieves all the goals. In presenting the process, we follow the structure of the rest of Section~\ref{sec:prep} \textit{viz.} we present how we address fixing, cleaning, adaptation, and scaling, in that order. Finally, as the states and sub-stages are many and intricate, this is a highly abridged version that includes the most important stages and illustrates the (nonobvious) complexity of the process. The full version appears in Appendix~\ref{app:cleaning}.

\minihead{Fixing} To fix \textbf{Library Versions}, we do a quick Internet
search for the nonexistent method. Most times
we found online discussion that suggested substitute methods
that appear in the version of \code{pandas} we use. Older versions of the
\code{pandas} documentation itself can also provide substitutes. In both cases, we
first tested whether the method is equivalent (or how we can make it), and then we changed the code accordingly. 

For the \textbf{Column Size Mismatch} issue, we checked if it was feasible to take a slice of the RHS
with the same size as the LHS column; if
this was impossible, we removed the code by following the
rules for cleaning.

Note that we don't have to fix \textit{every} error. In particular, if the error is caused by code that will be removed during cleaning, then we don't fix it.

\minihead{Cleaning} For cleaning we generally removed any import that is
unrelated to \code{pandas}, \code{numpy}, or the \code{Python} standard library
(most commonly, machine-learning and plotting libraries). To deal with the
\textbf{Dependencies} issue, we first identify manually all code that depends on non-Pandas
code. Then we delete it even if it includes Pandas code.

\minihead{Adaptation} For \textbf{Repeated Execution}, we delete any such code
that causes the notebook to run differently for subsequent runs. For \textbf{Lazy Evaluation}, we place function calls
strategically that evaluate the intermediate results (e.g., \code{evaluate\_eager(df.describe())}). Print calls cannot fulfill
this purpose because some libraries choose to display pre-evaluated contents
until lazy objects are explicitly evaluated. So, our evaluator functions need to
be more sophisticated. We could not find a single evaluator function that works
for \textit{all} the techniques, so we introduced one for each technique.
The code below shows the evaluator function we use for Dask:

\begin{minted}[bgcolor=light-gray]{python}
def evaluate_eager(lazy_obj):
  return lazy_obj.compute()
\end{minted}

\minihead{Data Scaling} We added a custom cell at the beginning of every
notebook that creates the scaled data, by creating new files, e.g., \code{input.scaled.csv}. We then replace all file
references in the rest of the notebook so that \code{input.csv} becomes
\code{input.scaled.csv}. This scaling had
to be done differently for every notebook as some notebooks used files with
special formats or encodings (e.g., a tab is used instead of comma as a
delimiter).

A final subtle step is to replace hardcoded values that depend on the size of
the input data (e.g., a list that is created with a hardcoded length \code{20},
but this list's function in the notebook depends on the data scale). We replace these
values with dynamic values.
\section{Exploring \system{}}
\label{sec:explore}

After finalizing the collection, cleaning, and fixing processes, we explored the
resulting collection of notebooks for interesting characteristics. In
particular, we explore the coverage of the benchmark (for a subset of the Pandas API), and also explore whether the input data captures what prior work~\cite{modin_first} observed makes Pandas unique.

\subsection{Coverage}

The overarching question we are concerned with in this section is: what percentage of the Pandas API is used by the benchmark?

We consider a subset of the Pandas API, which includes all the methods
and properties under \code{pandas.*}, \code{pandas.Series.*}, and
\code{pandas.DataFrame.*}. So, for example, our subset includes
\code{read\_csv()} but not anything under \code{pandas.api.plotting}. We gathered usage statistics dynamically because we find it to be much more
accurate and reliable than a static analysis. We only log top-level members, i.e., user-written code, because we believe it's more useful. For example, \code{df.head()} internally calls \code{iloc}. Even though \code{iloc} is in our subset, we do
not log it in this case.

The percentage of the subset API that is covered---meaning the percentage of
members that were invoked at least once---is: 32.9\%.

\subsection{Input Data}
\label{sec:expl:input}

We explore the data used by the notebooks, first for simple characteristics like
the data size, and then to see if they reach the boundary of what makes
dataframes different from other abstractions.

\begin{figure}[ht]
  \centering
  \includegraphics[width=1\columnwidth]{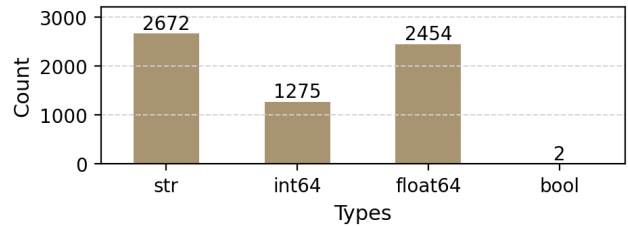}
  \caption{Histogram of column types of the datasets that are loaded in \system{}.}
  \label{plt:expl:types-count}
\end{figure}

\paragraph{\textbf{Histogram of Column Types}} First, we count how many times each column type appears, shown in
Figure~\ref{plt:expl:types-count}. In the whole benchmark, only 4 types are
used. \code{bool} is used much less than the other types. We also see that
\code{str} is used heavily, which is important because \code{str} is much slower
than the other three types because it can't be lowered to optimized native
operations.

Next, we want to check if any of the input data uses any features available only
to dataframes. Modin \cite{modin_first} identified several unique challenges of
the dataframe data model that don't exist e.g., in the relational model. We
focus on what we believe are the two more prominent; schema inference and
unbounded column cardinality.

\paragraph{\textbf{Schema Inference}}\label{para:schema-inf} In the Pandas API, the column types of a dataframe (the schema) are not
pre-defined (as e.g., in an RDBMS). Thus, any Pandas API implementation needs an inference engine
to determine the schema at runtime, by choosing the strongest possible type. For example, suppose we
read a CSV. In the input, all the data of a column is just a list of strings.
But, if all the data can be interpreted as integers (e.g., the strings contain
only numeric characters), then the column gets the type e.g., \code{int64} and
the column is converted into a numpy array under the hood (which allows much
faster processing).

We want to determine if any of the input data stresses an implementation's
inference engine. Conveniently, \code{pandas} tells us when this happens for
\textit{its} inference engine: it issues a warning that it's hard to determine
the type of a column because it has ``mixed types.''\footnote{An issue which can
easily be taken care of by using the \code{low\_memory=False} argument.} This is not the only way to stress the inference engine. Sometimes, we need to specify a
custom encoding (e.g., ISO-8859-1) to correctly read a file. Other times, we need to tell \code{read\_csv()} to
skip bad lines because they cannot be handled.\footnote{In both cases,
\code{pandas} will fail if the input is not appropriate, and that's how we
identify such instances.} So, we can find out all the input files of \system{} that stress the inference engine by just loading all of them with \code{pandas} and tracking the warnings. In total, the benchmark loads 11 files (\textasciitilde10\%), all of
them \code{.csv} or \code{.tsv}, that stress the inference engine.

\paragraph{\textbf{Unbounded Column Cardinality}} As Modin~\cite{modin_first}
identified, another unique feature of dataframes is that the number of columns
is ``unlimited.'' This is different from, say, conventional relational database
management systems (RDBMSs) which have an upper bound on the number of columns
(e.g., 1,600 in Postgres). Of course, a fair question is which dataset, and why,
contains more columns than this large number. Indeed, in our experience we have
not come across such a dataset (used by a Pandas notebook), and there is no such
dataset used by \system{}. Rather, such dataframes are almost always derivative,
and more specifically, a transposed version of an input dataset. This is because
many useful operations in Pandas (and \code{numpy}) are columnar (e.g.,
\code{np.corrcoef()}). A user may want to perform such an operation over the
rows. To achieve that, they transpose (i.e., \code{.T}) the dataframe, something
that: a) cannot be done in a RDBMS easily or at all, b) is extremely slow, and
c) usually hits a limit on the number of columns, as we mentioned earlier. As a
concrete example, a notebook~\cite{NB_deffro_eda-is-fun} of \system{}, uses
\code{np.corrcoef()} on a transposed DataFrame, which results in a DataFrame
with $4,446,966$ columns (which is way above the limit of most RDBMs, and this
is \textit{without} scaling).

\section{Evaluating Pandas Techniques}
\label{sec:eval}

In this section, we report experimental results of running a the set of Pandas
API optimization techniques over \system{}. Our goal is to demonstrate how
\system{} can be used to compare and evaluate a diverse set of techniques and to
identify opportunities for improvement. We use the following metrics: (a)
real-world coverage (\S\ref{sec:back-defs}), speedup (\S\ref{sec:eval-perf}),
and Memory \& disk usage (\S\ref{eval:mem-usg}). In this evaluation we refer to
notebooks by numerical index, which follows the lexicographic ordering of the
notebook names.

\subsection{Experimental Setup}

\subsubsection{Machine \& Python Environment}

All experiments were performed on a system with a dual-socket Intel Xeon 6348
Gold (28 cores each), 1 \emph{TB} of main memory, 6TB of local NVMe storage
(Dell PERC H755N Front), and Ubuntu 22.04.5 LTS. We used a \emph{single} machine
because \system{} targets single-machine workloads, but our machine is much more
powerful than a commodity machine so that the setup does not pose any obstacle.
This is especially true for our choice of main memory size, which is 1
\textit{terabyte}, to avoid potential memory issues given the high memory
consumption observed in prior work~\cite{dias}. We ran \system{} on Python
3.10.16. All library requirements are available in our source code.

\subsubsection{Baseline}

We use \code{pandas-1.5.1} as a baseline. \code{pandas} recently released version \code{2.x} which introduces major changes (e.g., the use of Arrow~\cite{arrow}), but not a lot of code is written with that version yet, and it is also not used in prior work~\cite{dias, modin_second}.

\subsubsection{Evaluated Pandas Optimization Techniques}

We first describe the Pandas API alternatives (\S\ref{sec:back-defs}) we used.
All Pandas alternatives we consider in this work share a conceptual framework
and technology. We did not choose them with that goal in mind; it is just the
case for the most used Pandas alternatives. In short, all of them aim to
mitigate the two main limitations of \code{pandas}: (1) it is single-threaded,
(2) the dataset has to fit into memory. The first is addressed by parallelizing
the API across cores and machines, while the second by using the disk when
necessary.

Modin~\cite{modin_first, modin_second} is possibly the most popular Pandas API
alternative, having been downloaded over 30 million times at the time of writing
\cite{modin_downloads}. Modin uses a parallel execution engine under the hood.
There are 2 which are currently supported: Ray~\cite{ray_paper,ray_arch}, which
is the default one---and thus the one we used---and Dask~\cite{dask}. Dask can
also be used independently, and it ``helps you process large tabular data by
parallelizing pandas, either on your laptop for larger-than-memory computing, or
on a distributed cluster of computers''~\cite{dask_website}. Finally, we also
consider Koalas~\cite{koalas}, which implements the Pandas API over
Spark~\cite{spark}.

Next, we describe the Pandas API complements we considered. Both radically different technology than the Pandas alternatives we just
described. On a high level, these techniques use program analysis and
transformation techniques.

Dias~\cite{dias} is a dynamic rewriter for Pandas code. It looks for, and
transforms, sub-optimal code using rewriting. The main benefit of Dias is that
it can optimize code that crosses library boundaries because it can access code
that a library doesn't. Dias has near 100\% coverage because if it fails to
match any code, the original code is executed; the only failure that can arise
are if Dias rewrites code incorrectly. We also considered SCRIPy~\cite{scirpy},
which also uses compiler technology (source-to-source transformations).
Unfortunately, the code of SCIRPy is not publicly available. We made a
best-effort attempt to contact the authors, but we were unsuccessful.

The versions of the techniques we use are: Modin\code{-0.17.0}, \\Dask\code{-2024.4.1},
Koalas\code{-3.5.1}, and Dias\code{-0.1.2}.

Finally, it's important to explicitly clarify why we do \textit{not} include
\textbf{PolaRS} in our evaluation, given that it is evaluated in prior
work~\cite{dias} and is one of the most popular Pandas replacements. The problem
is that PolaRS is \textit{not} a Pandas API optimization technique according to
our definition (\S\ref{sec:back-defs}), as it does not implement the Pandas
API~\cite{dias}. Thus, it is beyond the scope of this work.

\subsubsection{Technique Configurations}

For Pandas complements (\S\ref{sec:back-defs}), we need a base Pandas API
implementation that the techniques run on top of. We use \code{pandas} for this
purpose. For the Pandas alternatives, the most important knob is the number of
cores, which we set to something one could expect in environments like Kaggle
and Google Colab; specifically, to 4 CPU cores. For Koalas, some notebooks could
not run on Java 21 but could run on Java 17.

\subsubsection{Input Scales}
\label{sec:eval-scale}

As we explained in Section~\ref{sec:prep-scale}, we allow individual scale
factors. But we also need to understand what constitutes a good scale factor.
Our insight is that scaling factors should be computed based on a target running
time. In other words, we want a scaling factor such that \textit{each notebook
independently} runs for at least $X$ seconds. In particular, for a target
runtime of $X$ seconds, we scale the data of each notebook so that the notebook
runs with \code{pandas} for at least $X$ seconds, and more than $1.33 \times X$
seconds. 

\textit{For $X$, we used values of 5, 10, and 20 seconds}, which we use
throughout the evaluation. Each of them is a \textbf{target runtime}
(\textbf{TR}). In addition, we ran each Pandas API optimization technique with a
\textbf{default target runtime} where the data was not scaled. So, in total, we
have 4 target runtimes that we use throughout the evaluation.

Relative to the default target runtime, the other target runtimes will upscale
the input of some notebooks, but also \textit{downscale} it in others, to make
sure the upper bound of $1.33\times$ is met.  For example, a notebook that would
normally run for almost three minutes gets downscaled to run for a duration
similar to other notebooks in the target runtime. We place the upper bound so
that we can evaluate how increasing the scale affects \textit{all} notebooks
without excluding notebooks that run for much longer than the target time.

Now we describe how we compute the scale factors to meet the upper and lower bounds. To compute the scaling factor for a target runtime of $X$ seconds, we run each notebook starting with a factor of 1, and calculate the ratio between the target time and the current runtime. For notebooks with a ratio above 1, we increase their scale factor by slightly more than the calculated ratio. For ratios well below 1, we decrease their scale factor by the calculated ratio. We repeat until we converge on scale factors that give us runtimes within the bounds. This process was done automatically although in principle it could be automated.

\subsubsection{Measuring Memory \& Disk Usage} We will now
explain in detail how we measure memory consumption. Our method improves the accuracy
compared to prior work~\cite{dias}, but the difference does not change any of
the conclusions of that prior work.

To measure peak memory usage, after each cell runs we
programatically (using the \code{psutil} Python library.) measure the
Proportional Set Size of the main Python process, any
subprocesses a Pandas optimization technique may utilize (both Modin and Koalas
create worker processes), and we avoid over-counting any shared memory among these
processes. Furthermore, both Modin and Koalas may spill data to disk. For Modin, after each
cell runs we get the amount of data spilled using the \code{ray} CLI. For
Koalas, memory metrics are automatically tracked; after the notebook finishes
running we retrieve the amount of memory spilled via a Spark REST API. Nothing
additional has to be done for Dask or Dias. With Dask, we run it using the
default threaded scheduler, which does not create subprocesses or spill to disk.
Dias does not alter how the underlying Pandas API implementation manages memory
so we do not have to make any considerations for it.

\subsection{Real-World Coverage}
\label{sec:eval-cover}

\begin{figure}[ht]
  \begin{subfigure}{0.475\columnwidth}
     \includegraphics[width=\columnwidth]{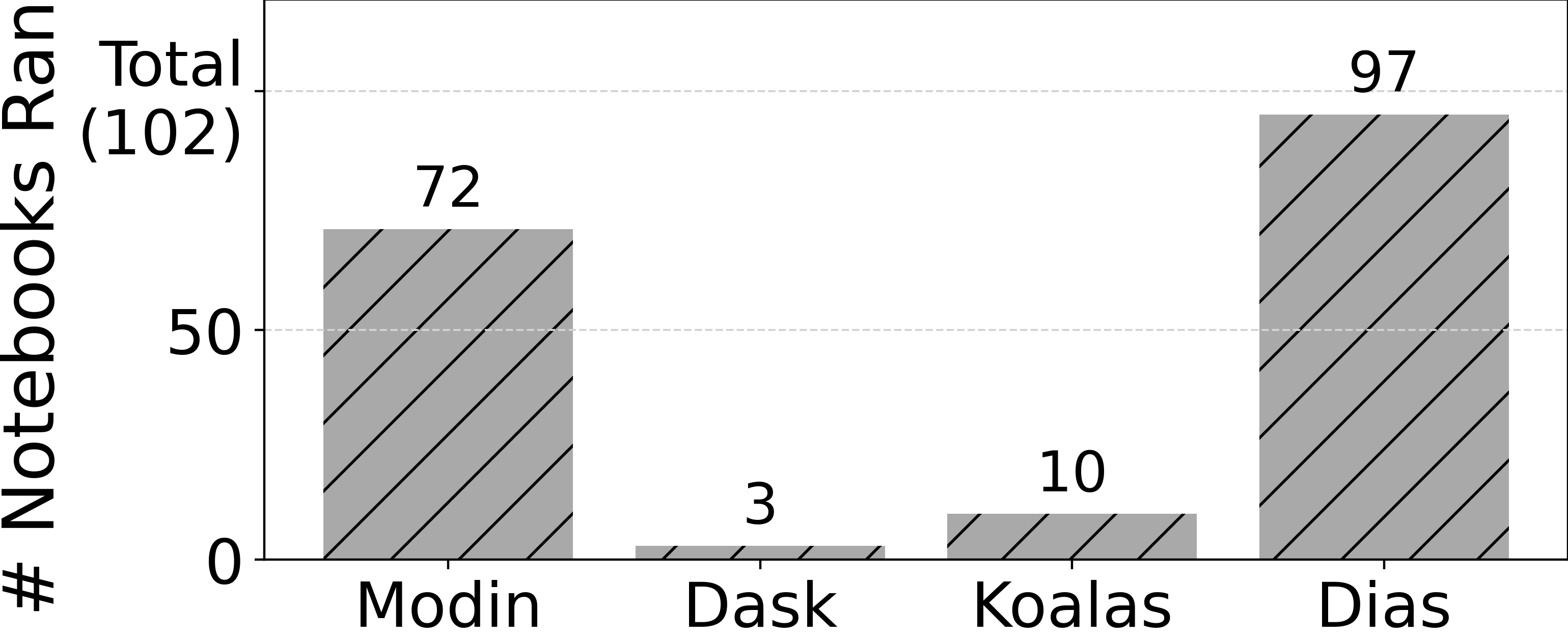}
    \caption{default target runtime}
    \label{subfig:num-nbs-ran-default}
  \end{subfigure}
  \hspace{2pt}
  \begin{subfigure}{0.475\columnwidth}
    \includegraphics[width=\columnwidth]{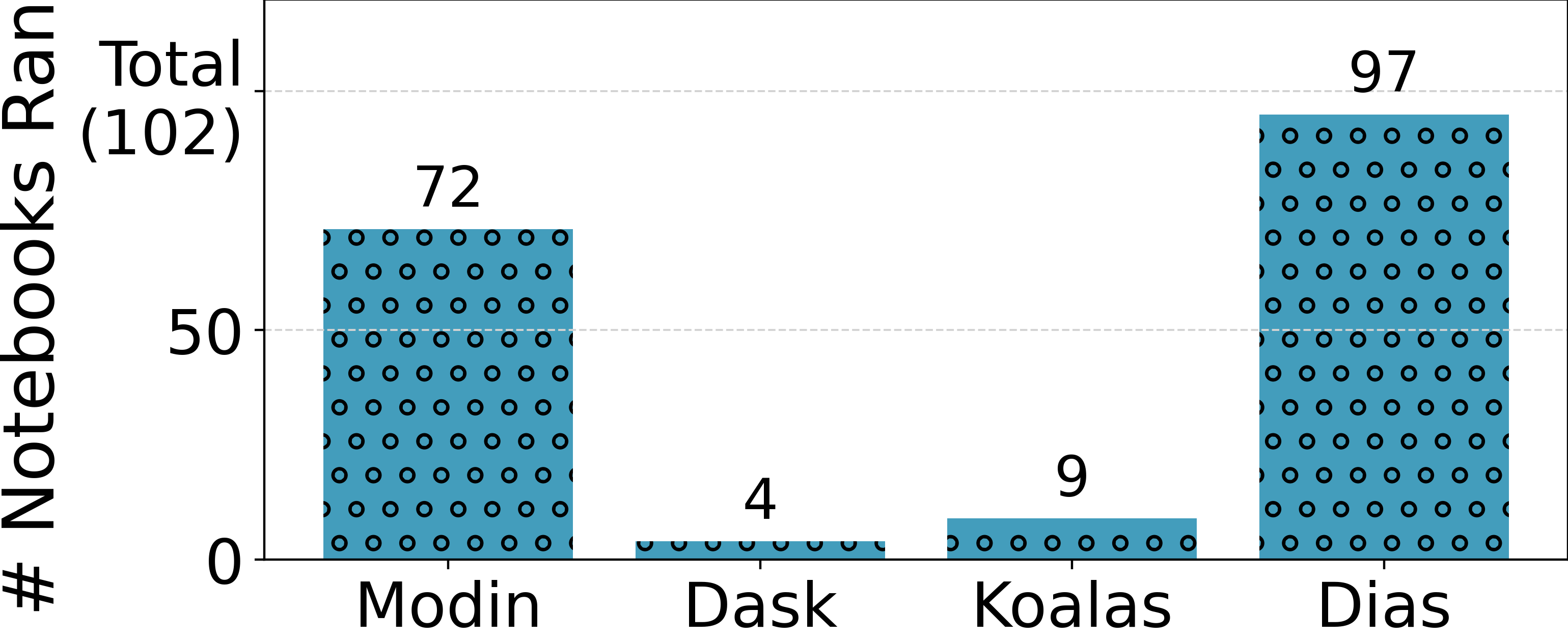}
    \caption{5-second target runtime}
    \label{subfig:num-nbs-ran-5sec}
  \end{subfigure}
  \begin{subfigure}{0.475\columnwidth}
     \includegraphics[width=\columnwidth]{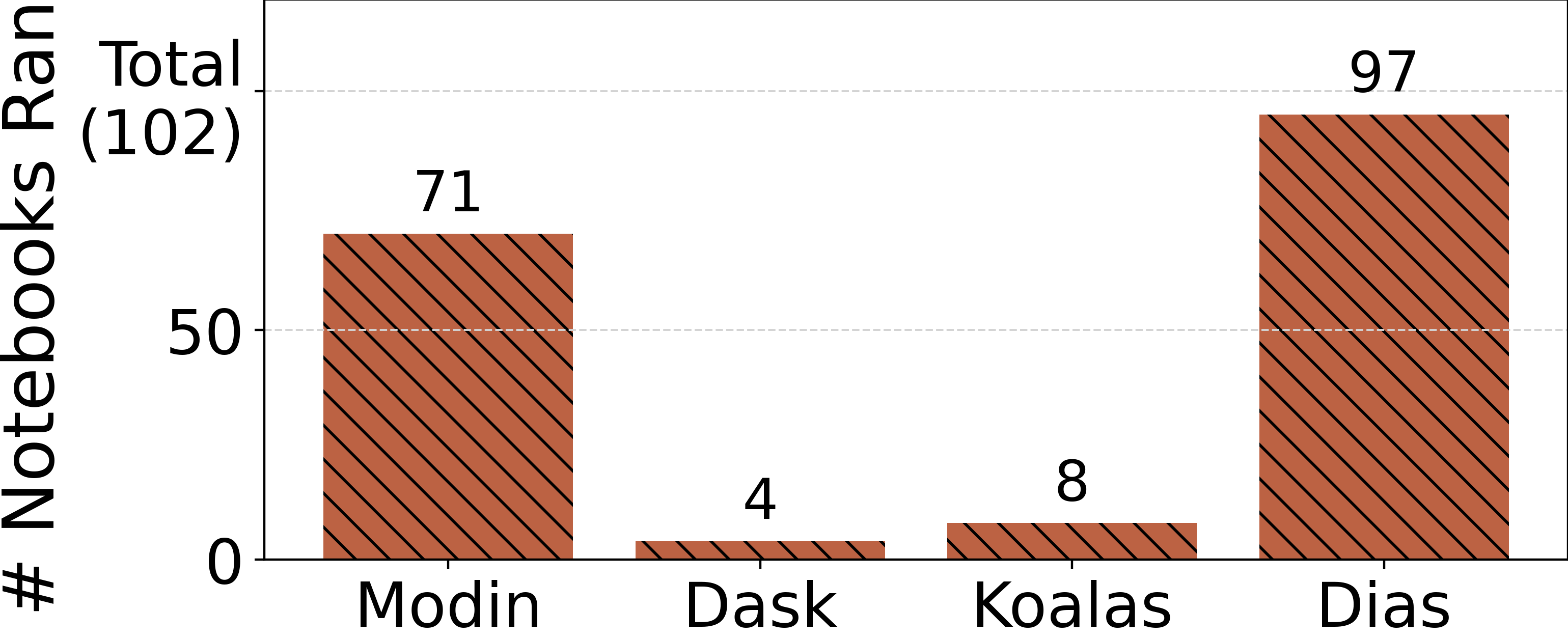}
    \caption{10-second target runtime}
    \label{subfig:num-nbs-ran-10sec}
  \end{subfigure}
  \hspace{2pt}
  \begin{subfigure}{0.475\columnwidth}
    \includegraphics[width=\columnwidth]{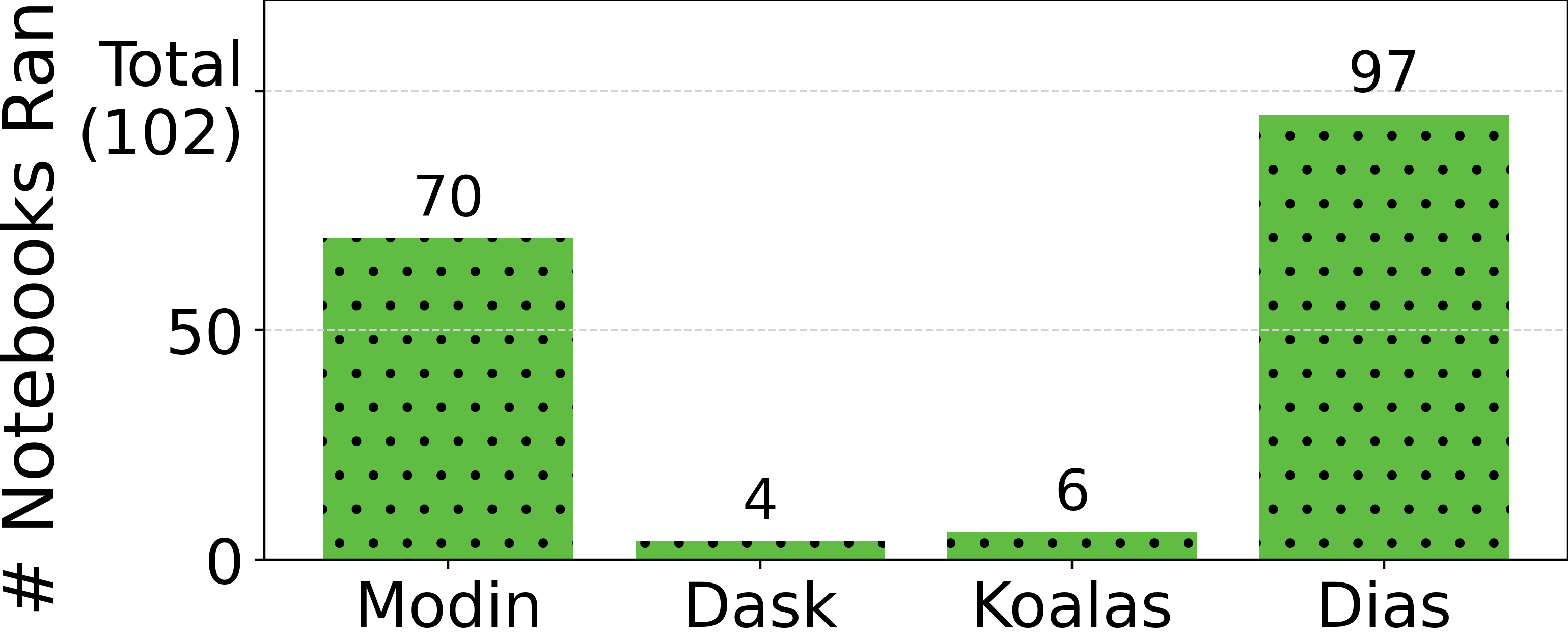}
    \caption{20-second target runtime}
    \label{subfig:num-nbs-ran-20sec}
  \end{subfigure}

  \caption{Number of notebooks that each Pandas optimization techniques ran across target runtimes.}
  \label{plt:num-nbs-ran}
\end{figure}

\renewcommand{\arraystretch}{1} 
\setlength{\tabcolsep}{12.7pt} 

\begin{table*}
    \centering
    \large
    \begin{tabular}{|c|c|c|c|c|c|c|}
        \hline
        \multirow{3}{*}{\makecell{\textbf{Pandas Optimization}\\\textbf{Technique}}} & \multicolumn{3}{c|}{Incompatible API} & \multirow{3}{*}{\makecell{Invalid\\Output}} & \multirow{3}{*}{\makecell{Bad Type\\Inference}} & \multirow{3}{*}{\makecell{Scaling\\Error}} \\
        \cline{2-4}
        & \makecell{Missing\\Method} & \makecell{Missing\\Argument} & \makecell{Unsupported\\Use Case} & & & \\
        \hline
        Modin & \mybullet & \mybullet & \mybullet & \mybullet & & \mybullet \\
        \hline
        Dask & \mybullet & \mybullet & \mybullet & & \mybullet & \\
        \hline
        Koalas & \mybullet & \mybullet & \mybullet & \mybullet & & \mybullet \\
        \hline
        Dias & & & & \mybullet & & \\ 
        \hline
    \end{tabular}
    \caption{Reasons for failure across techniques. A filled cell means that we observed this failure with this technique.}
    \label{tab:failure}
\end{table*}

\minihead{Overview} We first investigated which cells and notebooks of
\system{} each technique could run; that is, we examined the real-world coverage
of the techniques. 

We say that a Pandas optimization technique ran (i.e., did not fail to run) a
cell/notebook if no runtime errors occurred during its
execution (warnings were ignored). This implies that if a
\textit{single} cell fails to run, then the whole notebook is
considered to have failed. 

We present a notebook-level analysis and a cell-level analysis. First, we
present our findings and observations for the \textit{default} target runtime.
Because we observed a large number of failures in that, our next step was to
analyze why these took place. Finally, we tested what happens for different
target runtimes. We didn't expect any differences in coverage across target
runtimes (because only the input size changes), yet we did observe
discrepancies. Thus, at the end of this section we attempt to interpret the
impact of scaling.

\minihead{Notebook-Level Coverage} \system{} has a total of \numnbs{} notebooks.
We examined how many notebooks each technique ran in the default target runtime,
and present our results in Figure~\ref{subfig:num-nbs-ran-default}. Modin ran 72
out of \numnbs{} notebooks (\textasciitilde71\%). This is much higher than that
of Dask or Koalas, which could run only 3 (\textasciitilde3\%) and 10 notebooks
(\textasciitilde10\%), respectively.

Evidently, (real-world) coverage is surprisingly low across \textit{all} Pandas
\textit{alternatives} we consider, which seems to render these libraries unable
to be used as drop-in replacements for \code{pandas}.

Dias, on the other hand, achieves a much better coverage as it ran 97
(\textasciitilde95\%) of the \numnbs{} notebooks. However, this is expected
because Dias is merely a Pandas API \textit{complement}
(\S\ref{sec:back-defs}).

\minihead{Cell-Level Coverage} \system{} has a total of \numcells{} cells. We
examined how many of these cells each technique ran, focusing on the default
target runtime. Modin was able to run 2,870 cells (\textasciitilde77\%), Dask
was able to run 447 cells (\textasciitilde12\%), Koalas was able to run 844
cells (\textasciitilde23\%), and Dias was able to run 3,449 cells
(\textasciitilde93\%).

\subsubsection{Reasons For Failure}
\label{eval:reasons-fail}

We investigated what \textit{caused} the many failures we saw. We first present
cumulative results in Table~\ref{tab:failure}, and then we analyze individual
case studies. In particular, the the rows list the Pandas techniques and the
columns list all the reasons for failure we observed. This table shows which
reasons apply to which techniques. We will explain each reason separately, along
with examples, except for \emph{Scaling Error} which we will examine when we
consider the impact of scaling. 


\minihead{Incomplete API $\rightarrow$ Missing Method} Let us first consider the case of \emph{Incomplete API} issues, which Modin,
and Koalas, and Dask face. One failure case is that of a \emph{Missing
Method}, i.e., when a method is not implemented/supported. For example, both
Modin and Dask cannot run the following code~\cite{NB_nihhaar_etltransform} because of a missing method, but in
slightly different ways:

\begin{minted}[bgcolor=light-gray]{python}
pd.cut(df['# num_pages'], [-1, 100, 200, 300]).head()
\end{minted}

Modin does not implement \code{.head()} for the \code{Categorical} type returned
by \code{.cut()}, whereas Dask simply does not implement \code{.cut()}. Both
cause a failure, but this examination allows us to distinguish between them
qualitatively: it's better not to implement a method only under a certain
context (Modin), rather than not implementing a method at all (Dask).

\minihead{Incomplete API $\rightarrow$ Missing Parameter} In this case a method
\textit{is} supported, but some input parameters are not. Dask and Koalas
suffer heavily from this problem, and neither can run this code~\cite{NB_brianmendieta_data-cleaning-plus-eda}:

\begin{minted}[bgcolor=light-gray]{python}
job_data.drop('index',axis=1,inplace=True)
\end{minted}

This is because neither supports the \code{inplace} keyword argument. This
argument is ubiquitous: it appears 47 (46\%) notebooks, which therefore neither
Dask nor Koalas can run, and at least 8 Pandas API methods.\footnote{It is useful because it allows us to modify a
dataframe in place instead of creating a copy (which can save both memory
space and construction/copying overhead). Dias also identified the usefulness of
this keyword in the \code{InplaceUpdate} rewrite rule~\cite{dias}.}

However, similar to our discussion about methods earlier, the support of
arguments is also not a binary condition. That is, some keyword arguments are
supported but not fully by techniques. A case in point is \code{parse\_dates} in
\code{read\_csv()}. While Koalas accepts this argument, it only supports the
value \code{False}. Another example is of the \code{axis} argument in
\code{apply()}; Dask only supports the value 1. This is an important omission
because \code{apply()}---one of the main candidates for parallelization---can
only be applied across columns and not rows, which are usually far more than the
columns.\footnote{Also, one of the distinctive characteristics of the dataframe
model is the interchangeability of rows and columns~\cite{modin_downloads}.}

\minihead{Incomplete API $\rightarrow$ Unsupported Use Case} Finally, this
occurs when a Pandas optimization technique throws an error when an API
componenent is called on specific forms of an object (usually a dataframe) that:
a) are supported by \code{pandas}, and b) the documentation of the technique
does not mention that it does not support them. For instance, Koalas usually
throws an error when calling \code{transpose()} on a dataframe that has string
columns. Also, Koalas does not interoperate well with \code{numpy} types. For
example, suppose we use the (popular) \code{dtype} argument in
\code{read\_csv()} to explicitly specify the types of columns (and thus bypass
the inference engine). If we provide numpy types (e.g., \code{np.uint8}), then
Koalas fails.\footnote{\code{pandas} columns are \code{numpy} arrays under the
hood, so the interoperability is natural. Modin also seems to use \code{numpy}
types under the hood. Koalas (and PySpark), on the other hand, is based on a
custom integer type and not the \code{np.*} family.} A third example is shown
below~\cite{NB_jagangupta_stop-the-s-toxic-comments-eda}:
\begin{minted}[bgcolor=light-gray,escapeinside=||]{python}
print("perc   :",round(nrow_train*100/|sum|),
          "   :",round(nrow_test*100/|sum|))
\end{minted}

Here, lazily-evaluated expressions (the second and fourth arguments) are passed
as arguments to an eagerly-evaluated function (\code{print()}). We observe that
even though both Koalas and Dask use lazy evaluation, only the latter fails on
this snippet.

\minihead{Invalid Output} This is a \textit{correctness issue}. It is a failure that
was caused by an (earlier) incorrect result. We identify these by manual
debugging and investigation of the error/exception we get. For example, consider
the following code:
\begin{minted}[bgcolor=light-gray]{python}
df_products['product_color']
    = df_products['product_color'].str.split(' ')
\end{minted}

\noindent Dias rewrites it to to the following,  which calls \code{split()} on a
column with floats, when \code{split()} can only be used on strings.
\begin{minted}[bgcolor=light-gray]{python}
...
_REWR_ls = df_products['product_color'].tolist()
for _REWR_s in _REWR_ls:
  _REWR_spl = _REWR_s.split(' ')
  ...
\end{minted}

\minihead{Bad Type Inference} This issue happens only with Dask. The inference engine for Dask, used in \code{read\_csv}, may
fail if the user has not provided explicit column types. For instance, a
notebook~\cite{NB_ukveteran_sarima-time-series-eda-predict-jma-sao-paulo}
contains a \code{read\_csv()}, and Dask infers that a column is
an integer type, only to later find floats in said column, causing
\code{read\_csv()} to fail.\footnote{The original input has 36 rows, some of which
are missing data for a column names \code{'obitos por dia'}. Dask infers the
type of \code{'obitos por dia'} as integer and throws an error upon
encountering the missing values, which it interprets as floats.} This highlights the importance of a robust
inference engine that matches how \code{pandas} handles
such edge cases (\S\ref{para:schema-inf}).

\minihead{Impact of Scaling} If we take a look at Figure~\ref{plt:num-nbs-ran} we observe that as
we scale the input, \textit{fewer} cells (and so, notebooks) run. This is
surprising because, as is usual when one scales data (e.g., in TPC-H), our
scaling does not change anything apart from the \textit{size} of the data.
Namely, it does \textit{not} change any code or any data types; it does not even
change the types that any correct inference engine would infer. Thus, we
investigated this behavior further.

First, Table~\ref{tab:failure} includes \emph{Scaling Error} as a cause of
failure to indicate which techniques it applies to. If we look again at
Figure~\ref{plt:num-nbs-ran}, we can discern that the effect is especially
prevalent in Koalas. In particular, only 6 out of 10 (60\%) of the notebooks
that ran on the default target runtime also ran on the 20-second target runtime. 

Koalas fails for higher target runtimes primarily due to out of memory errors.
More specifically, Koalas runs on top of Spark~\cite{spark}, which itself runs
on top of JVM, using multiple processes (this is how Koalas can be
multi-threaded). These JVM processes run out of memory for larger target
runtimes. We remind the reader that we ran Koalas out of the box in a machine
with 1 \textit{terabyte} of main memory. We even attempted to fix the errors by
trying different settings, similar to (the artifact of) Dias~\cite{dias}, to no
avail.

The main reason this result was surprising is that Koalas (and similar Pandas alternatives) are supposed to work \textit{better} as the input size increases (something which in fact we observe this later in Section~\ref{eval:mem-usg} when it comes to memory consumption).


Modin also suffers from \emph{Scaling Error} but to a lesser extent than
Koalas. Unlike Koalas, Modin fails to run notebooks on higher target runtimes
not due to resource management issues, but due to how it internally processes
data. Here is an example~\cite{NB_guyalmog-hr-analytics-predicting-employees-attrition}:
\begin{minted}[bgcolor=light-gray]{python}
x.Age = pd.cut(x.Age, 4)
\end{minted}
Modin fails with scaled data because it internally does not convert a
\code{Series} to a \code{numpy.ndarray} on larger target runtimes, and so when a
\code{numpy} method is called, an exception is raised.

Dask was another surprising case because it runs one \textit{more} notebook in the default TR than the 5s TR. This is because this notebook was \textit{down}scaled in the 5s (\S\ref{sec:eval-scale}). With the downscaling, a float that appears later in a column that Dask infers as int gets cut out, which ``avoids'' the inference error we mentioned in \textbf{Bad Type Inference}.

\subsection{Runtime Performance}
\label{sec:eval-perf}

In this section we examine the runtime performance of the various techniques,
focusing mainly on the speedup relative to \code{pandas}. We will also consider
memory usage at the end of the section. We exclude from our presentation
notebooks and cells that did \textit{not run at all}, as we explained in
\S\ref{sec:eval-cover}. This is, for example, why the x-axes in the plots do
\textit{not} include all \numnbs{} notebooks. All these x-axes of relative speedups are sorted based on the speedups (Figures 4-7).


\begin{figure*}[ht]
  \begin{subfigure}{1.045\columnwidth}
     \includegraphics[width=\columnwidth]{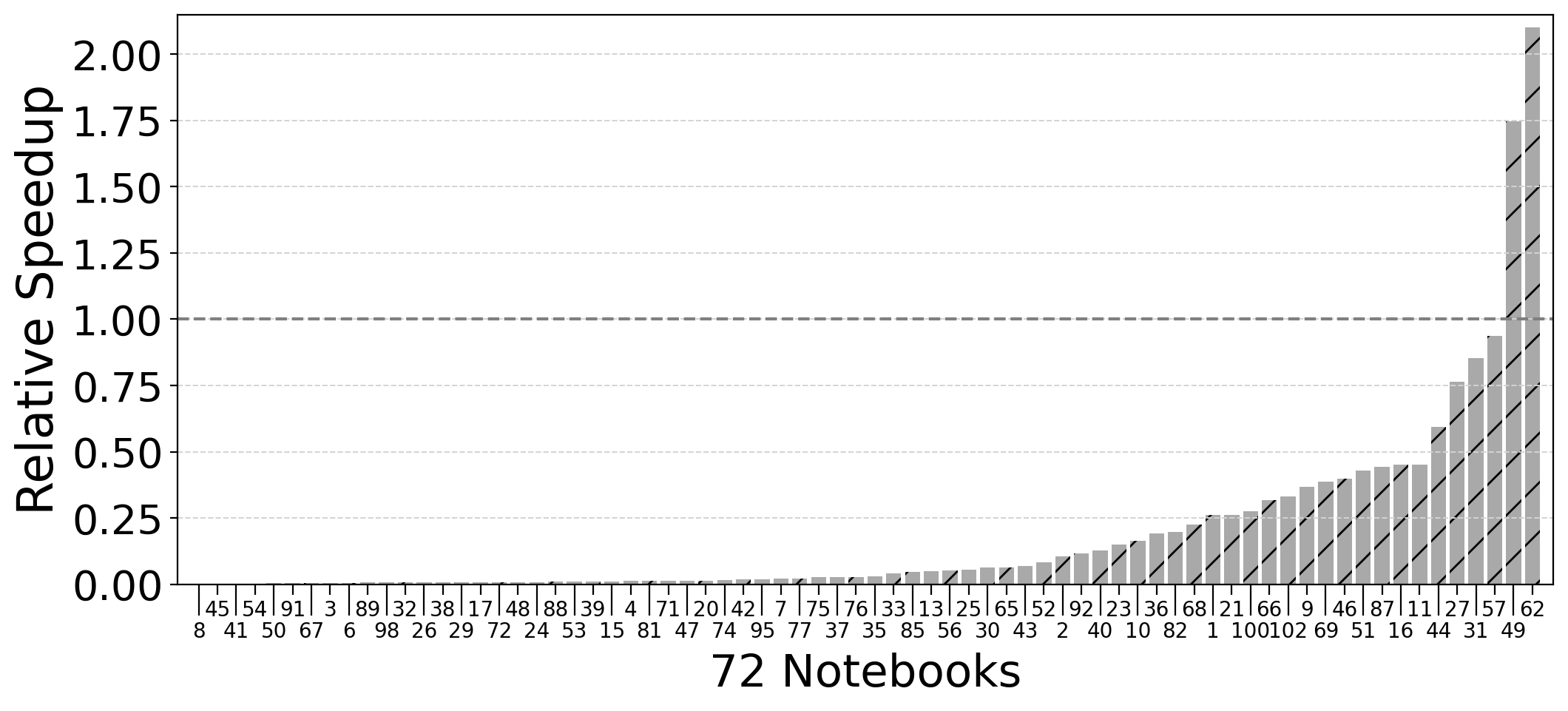}
    \caption{default target runtime}
    \label{subfig:modin-default}
  \end{subfigure}
  \begin{subfigure}{1.045\columnwidth}
    \includegraphics[width=\columnwidth]{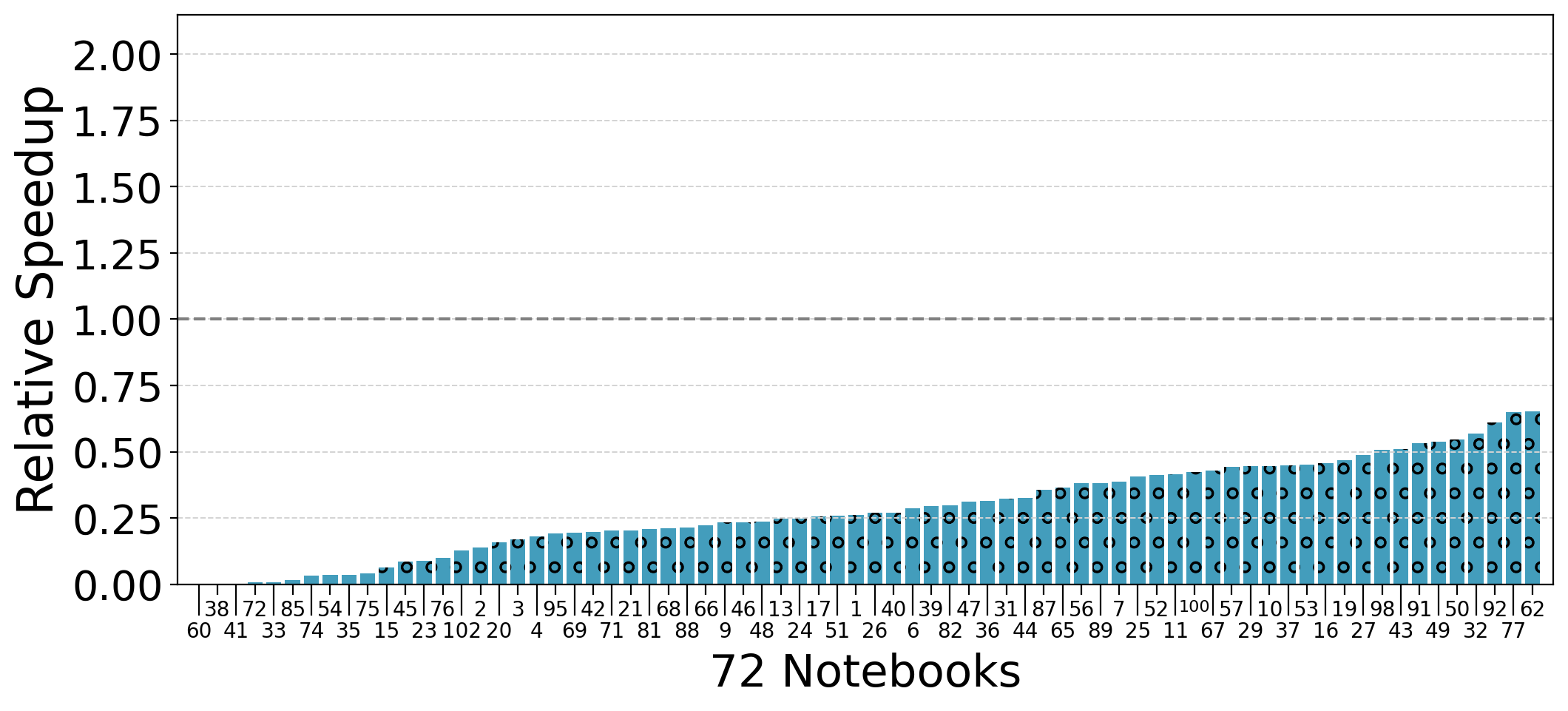}
    \caption{5-second target runtime.}
    \label{subfig:modin-5sec}
  \end{subfigure}
  \begin{subfigure}{1.045\columnwidth}
     \includegraphics[width=\columnwidth]{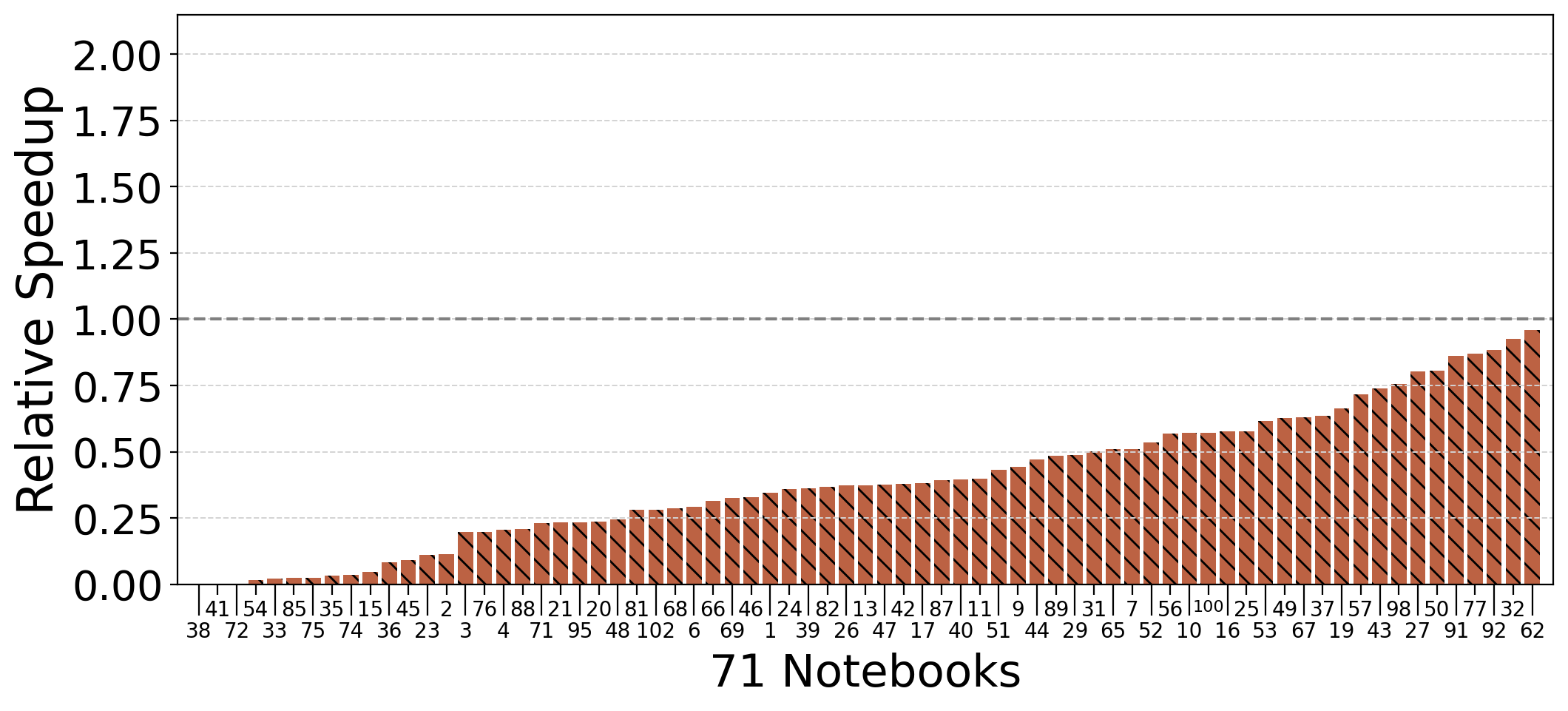}
    \caption{10-second target runtime}
    \label{subfig:modin-10sec}
  \end{subfigure}
  \begin{subfigure}{1.045\columnwidth}
     \includegraphics[width=\columnwidth]{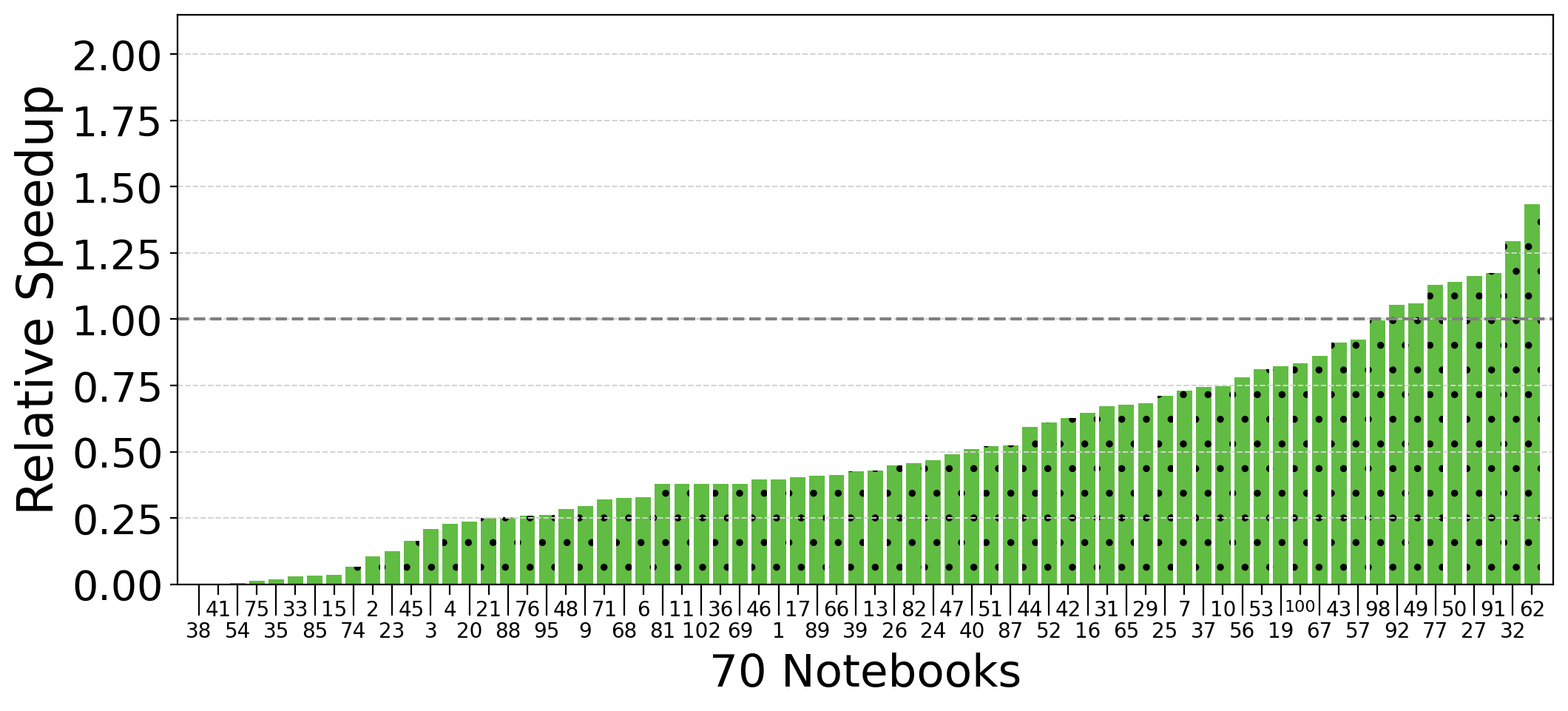}
    \caption{20-second target runtime}
    \label{subfig:modin-20sec}
  \end{subfigure}

  \caption{Per-notebook speedups of Modin compared to \code{pandas} across target runtimes. As we increase the scale, we see higher speedups, except for the default runtime.}
  \label{plt:speed-rel-modin}
\end{figure*}

\begin{figure}[ht]
  \begin{subfigure}{0.475\columnwidth}
     \includegraphics[width=\columnwidth]{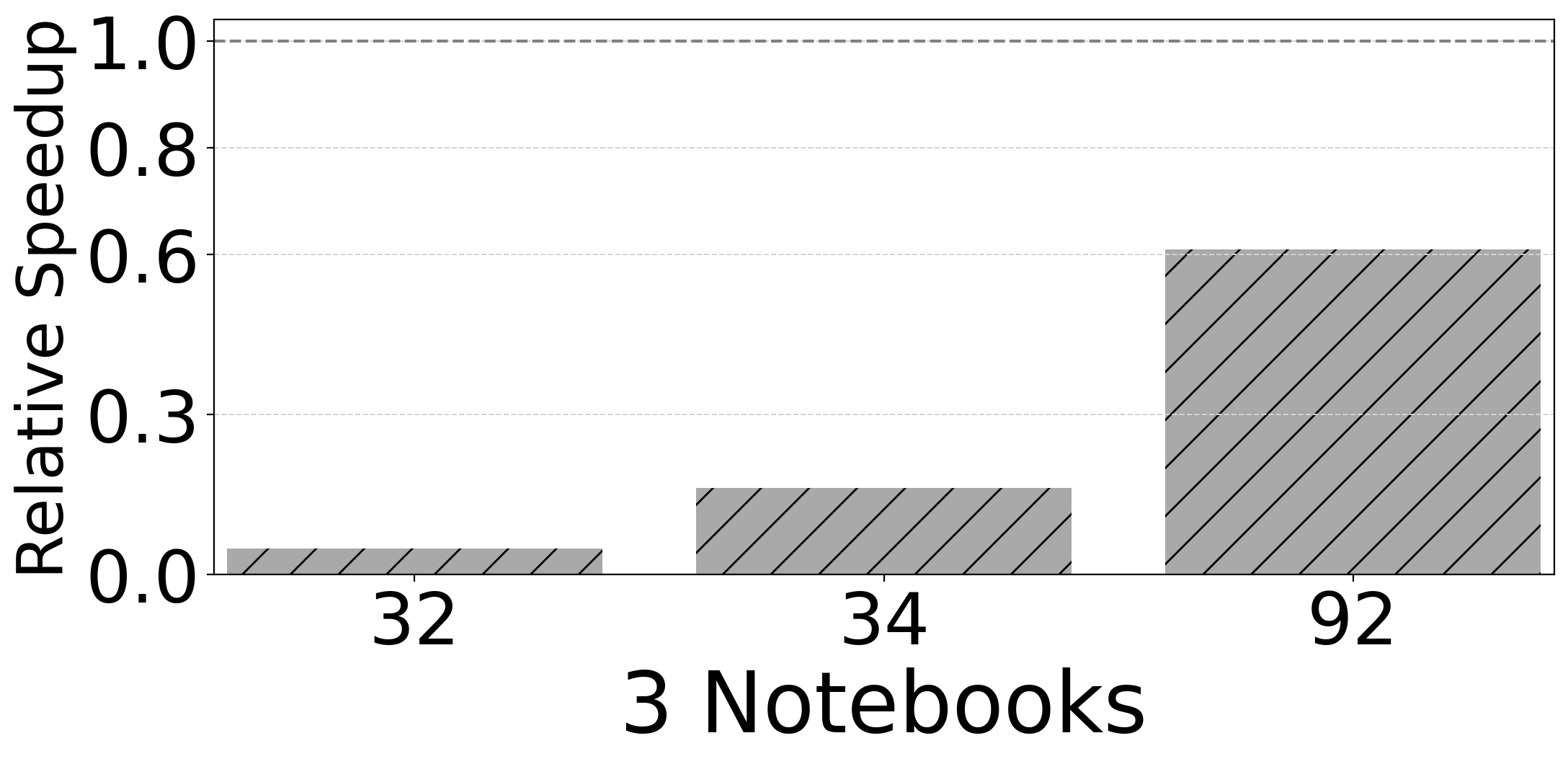}
    \caption{default runtime}
    \label{subfig:dask-default}
  \end{subfigure}
  \begin{subfigure}{0.475\columnwidth}
    \includegraphics[width=\columnwidth]{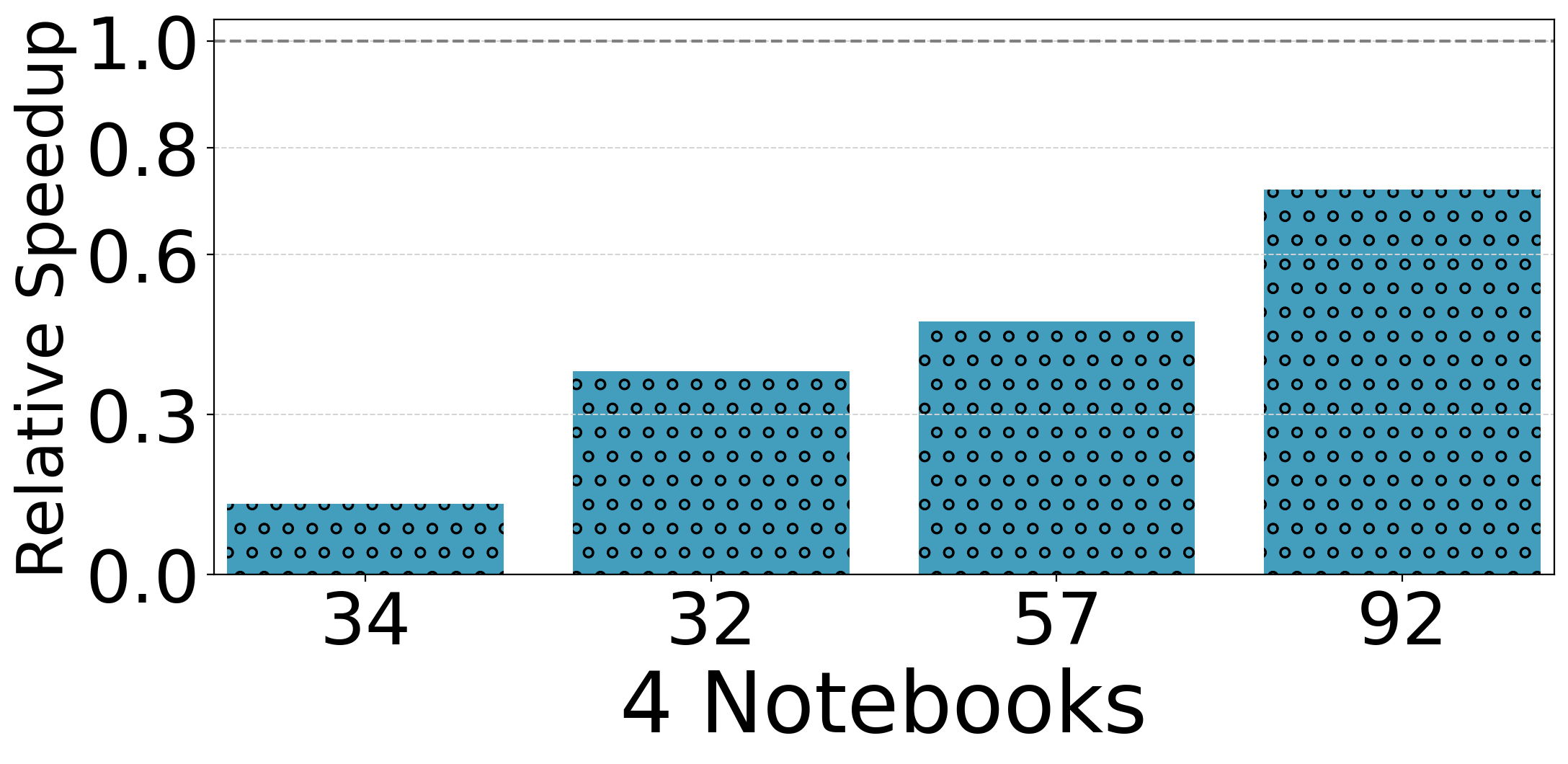}
    \caption{5-second target runtime}
    \label{subfig:dask-5sec}
  \end{subfigure}
  \begin{subfigure}{0.475\columnwidth}
     \includegraphics[width=\columnwidth]{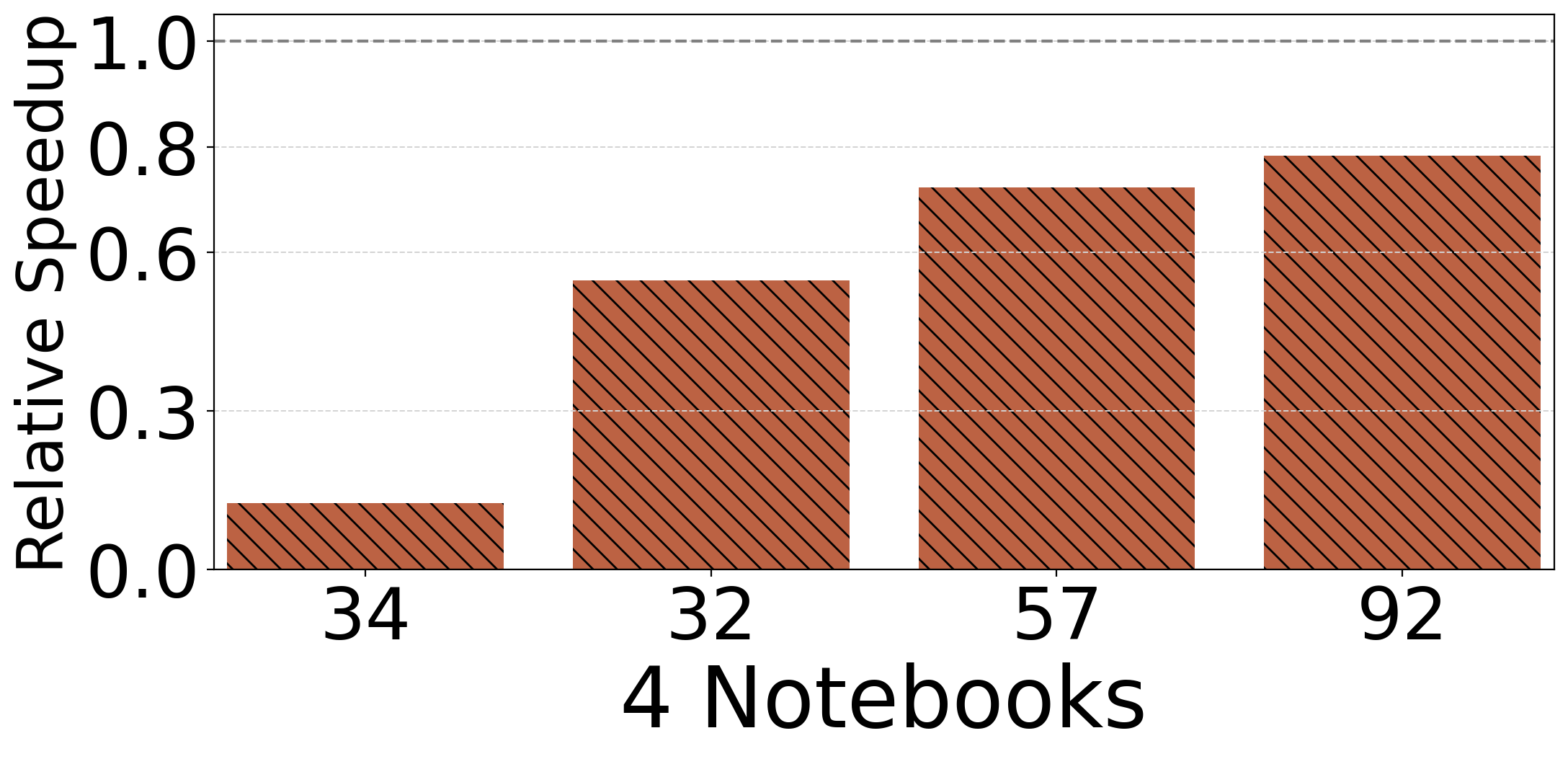}
    \caption{10-second target runtime}
    \label{subfig:dask-10sec}
  \end{subfigure}
  \begin{subfigure}{0.475\columnwidth}
    \includegraphics[width=\columnwidth]{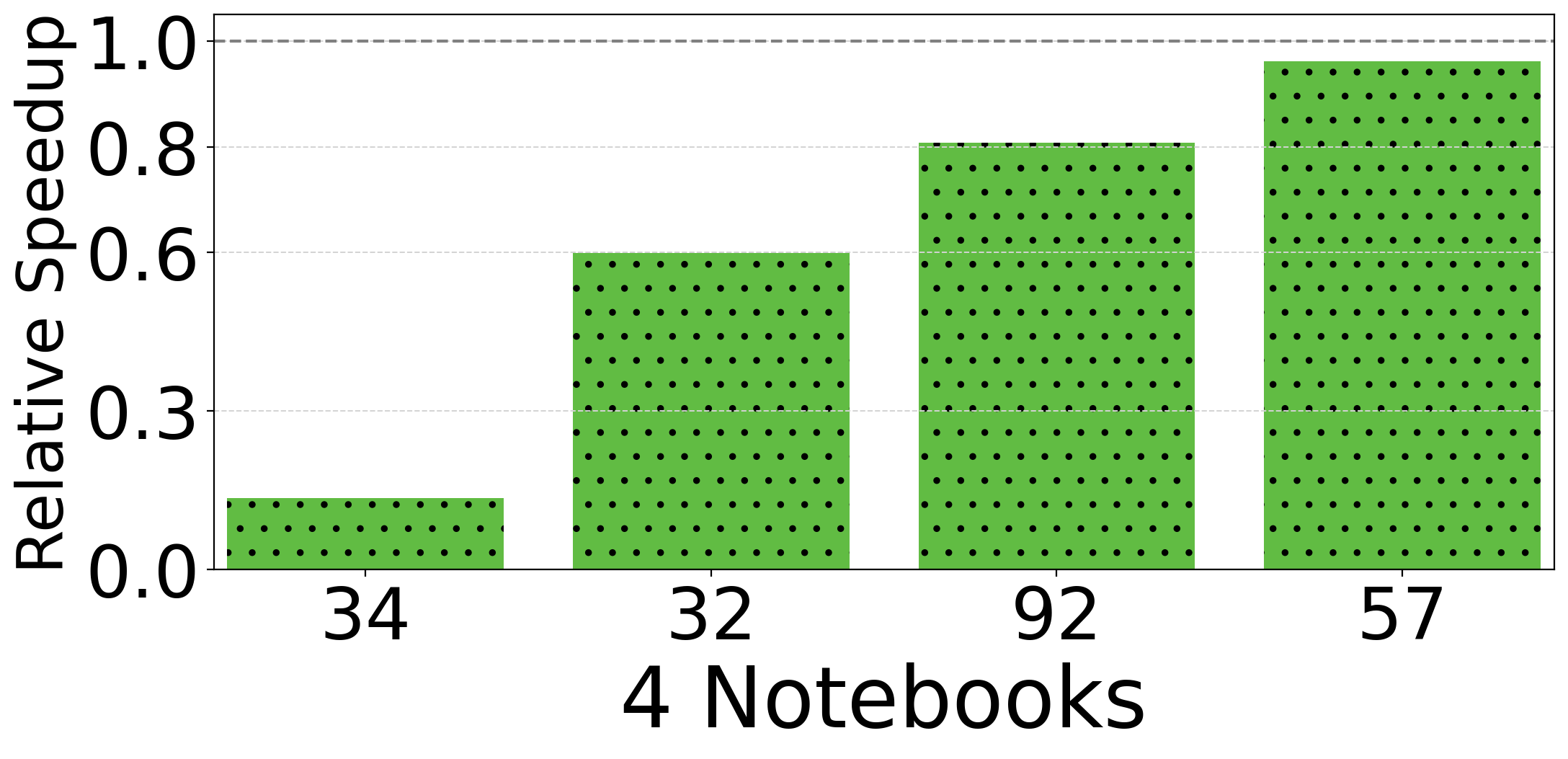}
    \caption{20-second target runtime}
    \label{subfig:dask-20sec}
  \end{subfigure}

  \caption{Per-notebook speedups of Dask compared to \code{pandas} across target runtimes.}
  \label{plt:speed-rel-dask}
\end{figure}

\begin{figure}[ht]
  \begin{subfigure}{0.475\columnwidth}
     \includegraphics[width=\columnwidth]{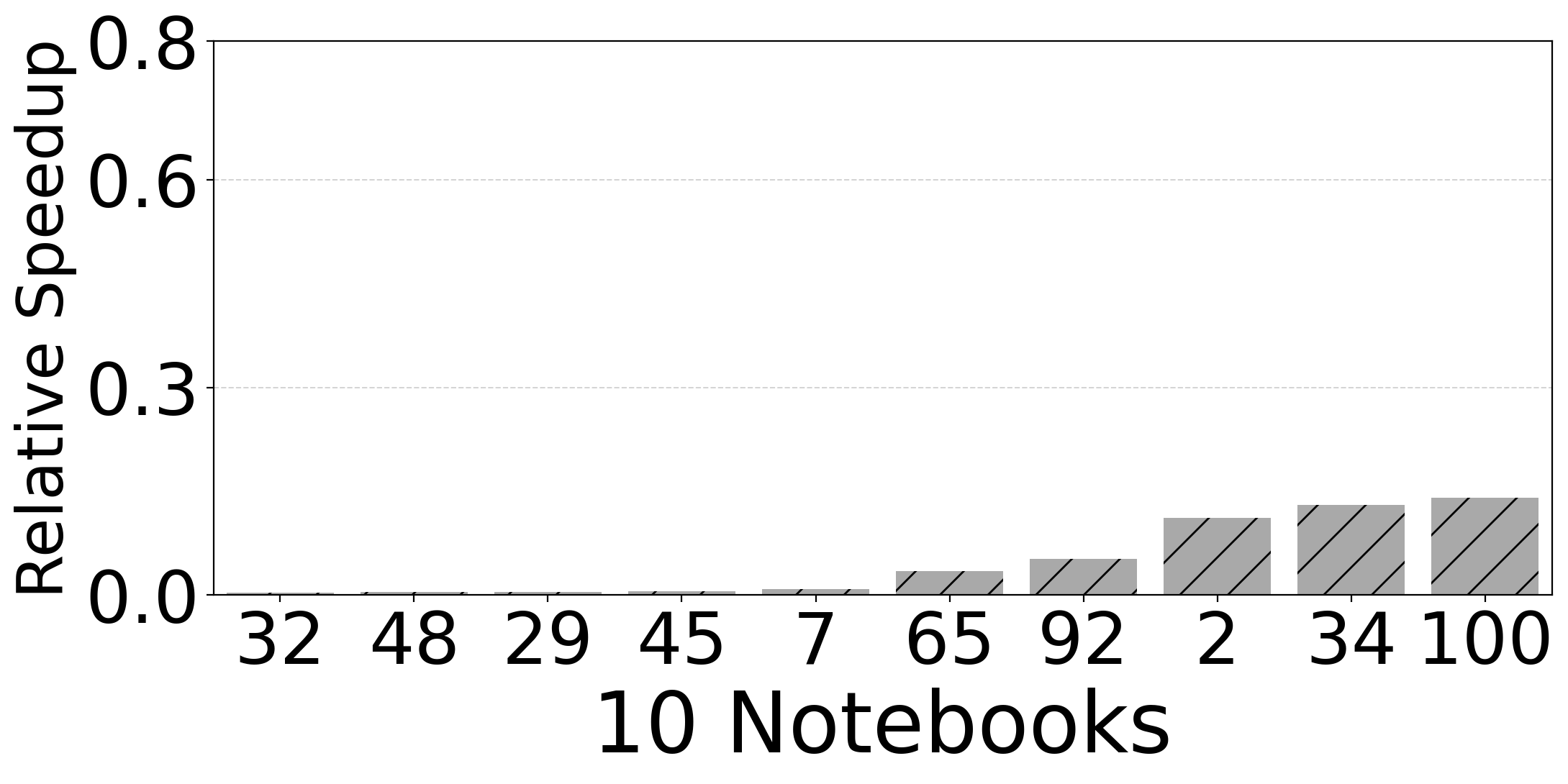}
    \caption{default target runtime}
    \label{subfig:koalas-default}
  \end{subfigure}
  \begin{subfigure}{0.475\columnwidth}
    \includegraphics[width=\columnwidth]{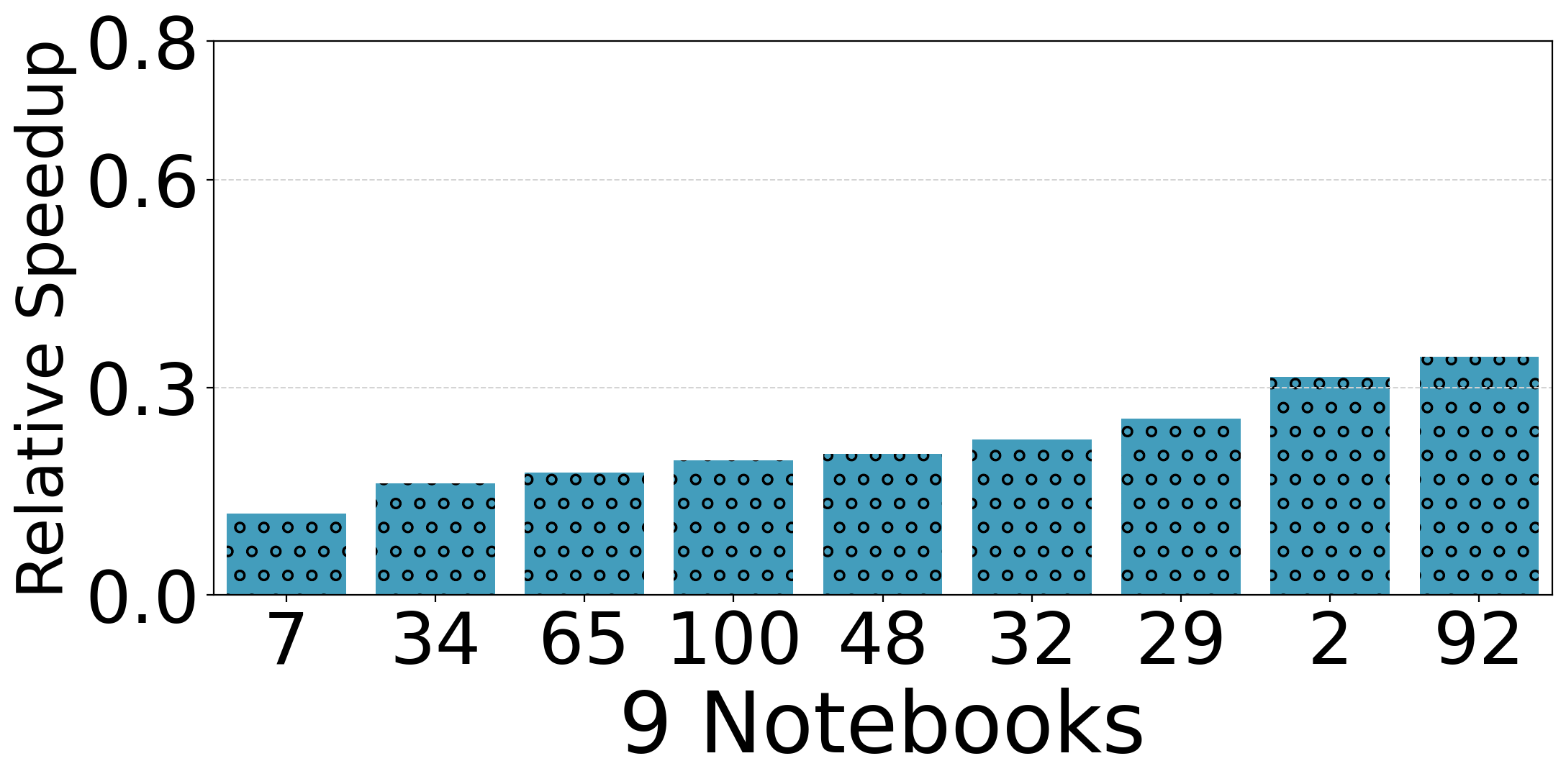}
    \caption{5-second target runtime}
    \label{subfig:koalas-5sec}
  \end{subfigure}
  \begin{subfigure}{0.475\columnwidth}
     \includegraphics[width=\columnwidth]{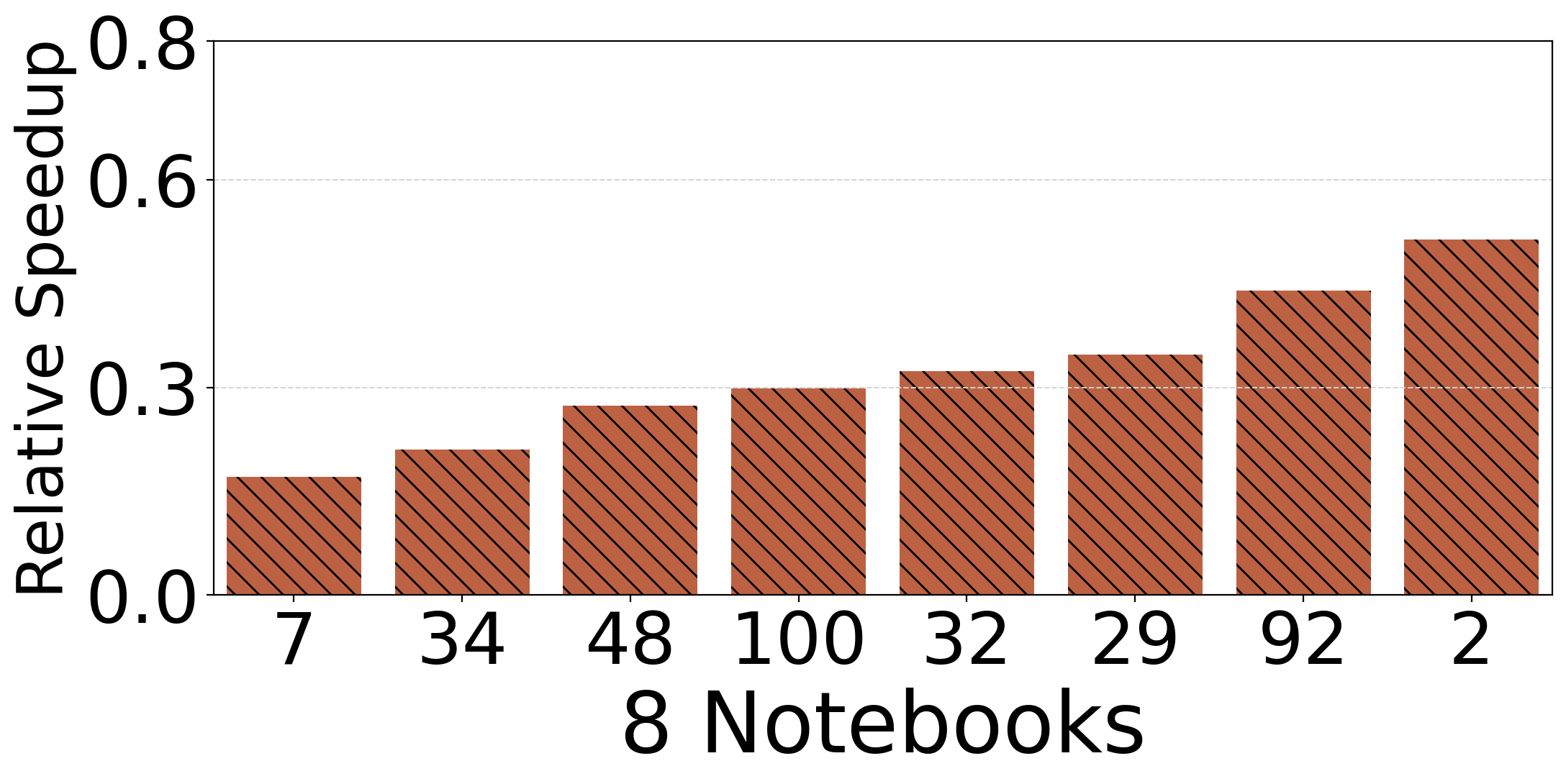}
    \caption{10-second target runtime}
    \label{subfig:koalas-10sec}
  \end{subfigure}
  \begin{subfigure}{0.475\columnwidth}
    \includegraphics[width=\columnwidth]{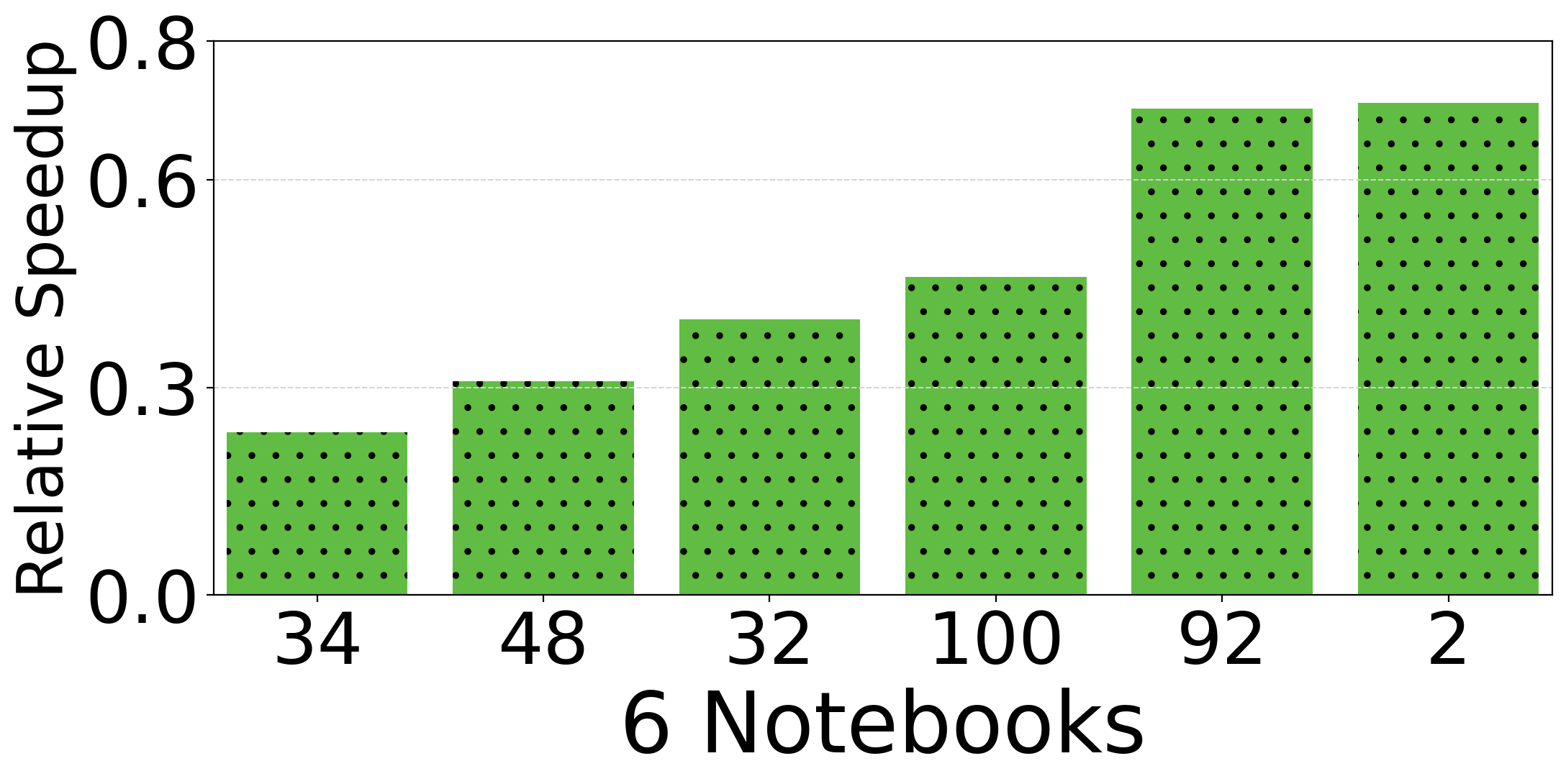}
    \caption{20-second target runtime}
    \label{subfig:koalas-20sec}
  \end{subfigure}

  \caption{Per-notebook speedups of Koalas compared to \code{pandas} across target runtimes. Koalas improves on larger target runtimes, getting closest to surpassing Pandas when run on the 20 second target runtime.}
  \label{plt:speed-rel-koalas}
\end{figure}

\begin{figure*}[ht]
  \begin{subfigure}{1.045\columnwidth}
     \includegraphics[width=\columnwidth]{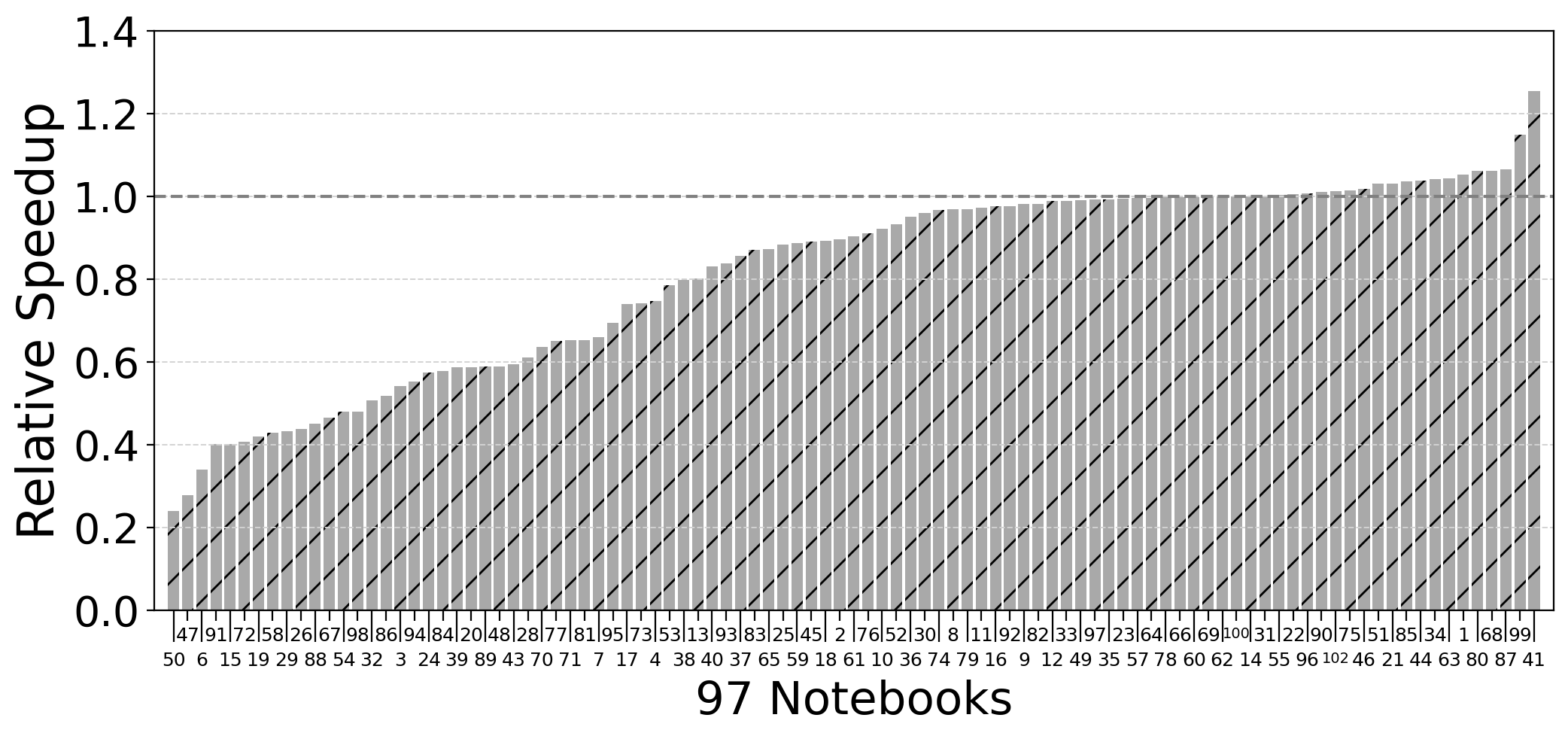}
    \caption{default target runtime}
    \label{subfig:dias-default}
  \end{subfigure}
  \begin{subfigure}{1.045\columnwidth}
    \includegraphics[width=\columnwidth]{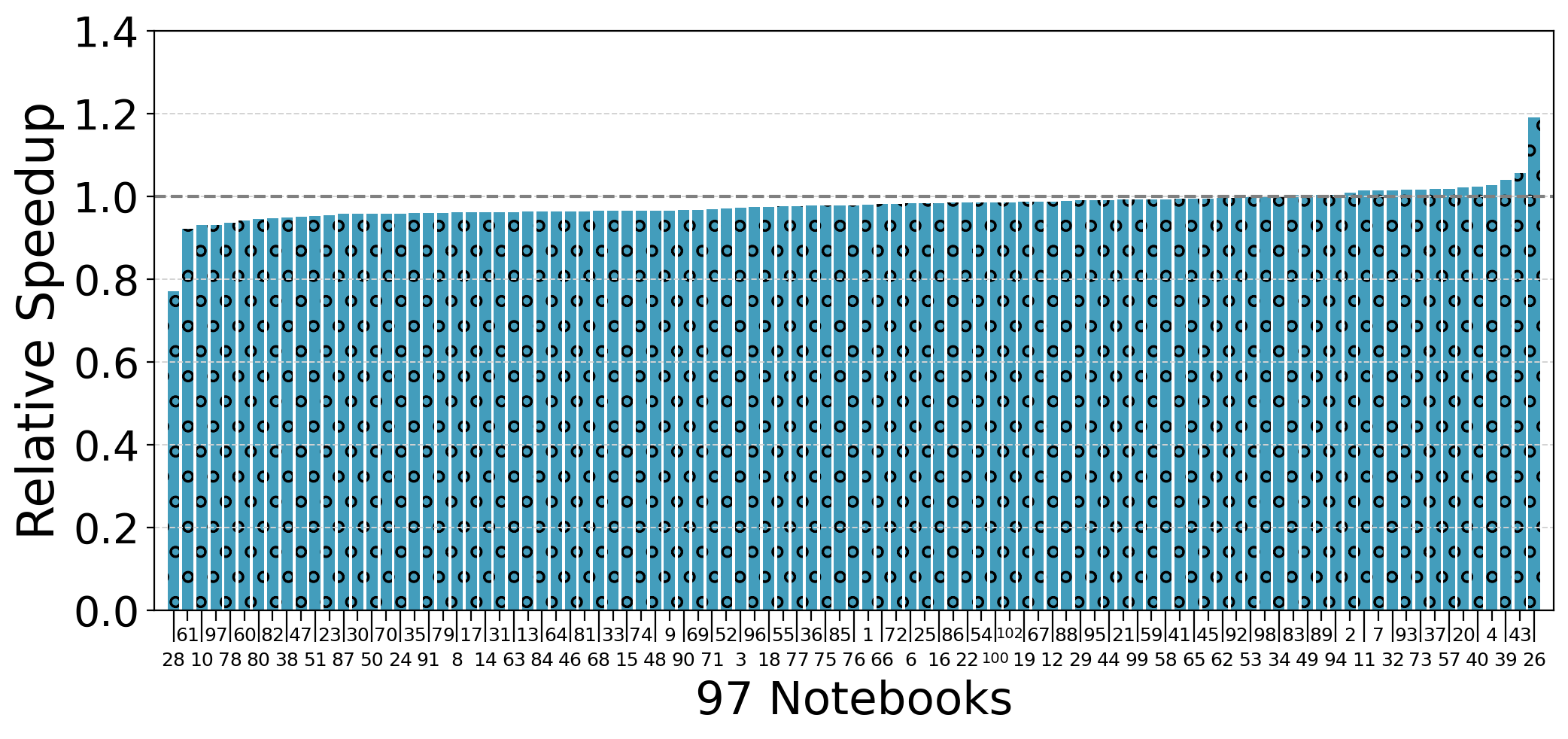}
    \caption{5-second target runtime}
    \label{subfig:dias-5sec}
  \end{subfigure}
  \begin{subfigure}{1.045\columnwidth}
     \includegraphics[width=\columnwidth]{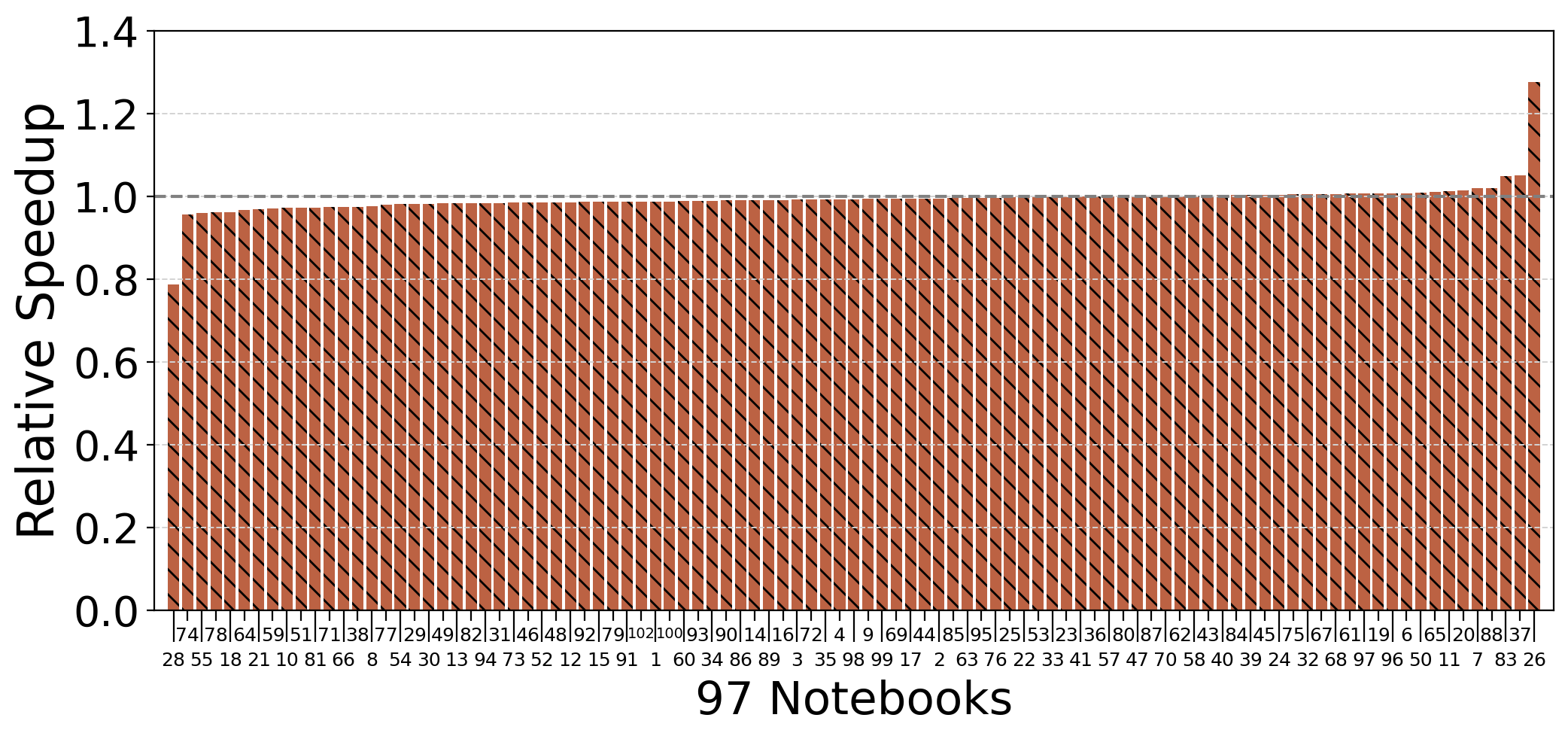}
    \caption{10-second target runtime}
    \label{subfig:dias-10sec}
  \end{subfigure}
  \begin{subfigure}{1.045\columnwidth}
    \includegraphics[width=\columnwidth]{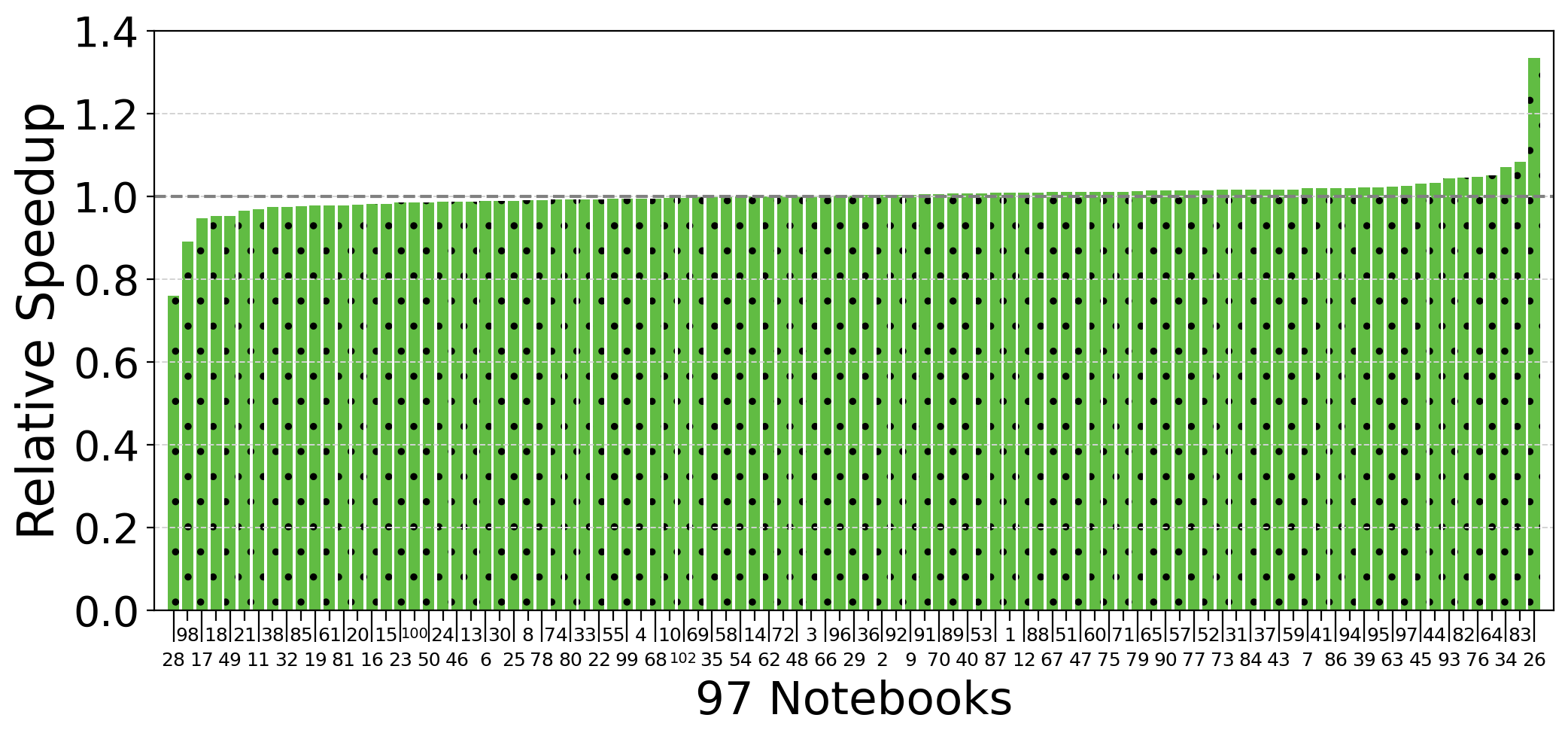}
    \caption{20-second target runtime}
    \label{subfig:dias-20sec}
  \end{subfigure}

  \caption{Per-notebook speedups of Dias compared to vanilla \code{pandas} across target runtimes. Dias runs most notebooks slightly slower than Pandas alone, speeds up a few notebooks, and significantly speeds up one notebook by almost 1.4x.}
  \label{plt:speed-rel-dias}
\end{figure*}
\begin{figure}[ht]
  \centering
  \includegraphics[width=1\columnwidth]{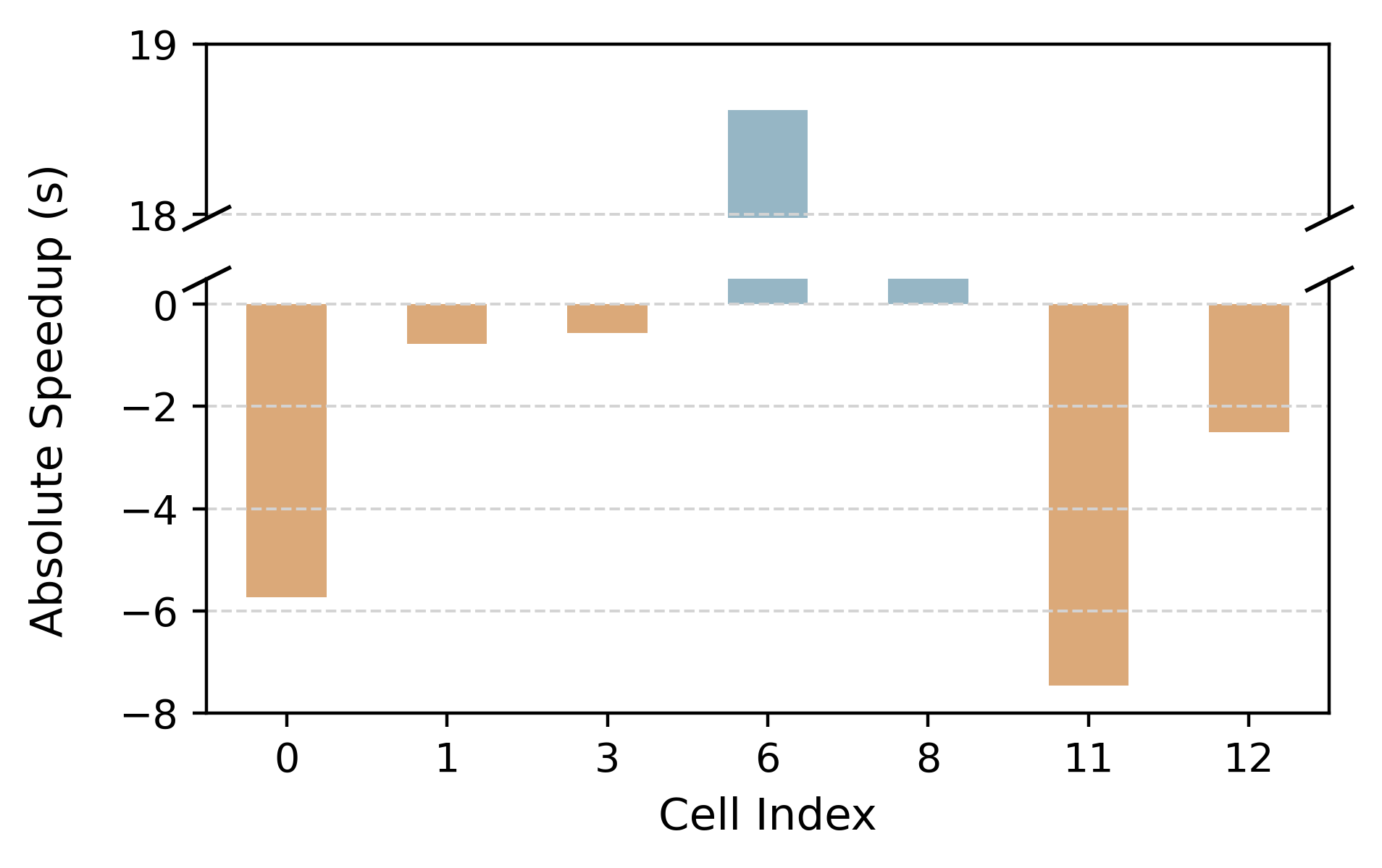}
  \caption{Per-cell absolute speedups (excluding speedups near 0) of the
  notebook~\cite{NB_poiss0nriot_classifying-client-type-using-client-names}
  Modin ran the fastest on the 20s target runtime. Modin takes almost 6
  seconds longer than \code{pandas} to run the first cell, but is faster than \code{pandas} by
  almost 19 seconds on cell 6.}
  \label{plt:per-cell-speedup-modin}
\end{figure}

\minihead{Modin} We first present results for Modin in Figure~\ref{plt:speed-rel-modin}. We observe that Modin \textit{slows down} most notebooks on all target runtimes. In the default target runtime (Figure~\ref{subfig:modin-default}), only 2 notebooks get a speedup, with the highest speedup in notebook 62, which is also the  highest in the 20s target runtime. Only these two target runtimes get any speedups at all. The 20s target gets the most speedups, with 8 notebooks in total. Still, \textasciitilde92\% get a slowdown. In the 5s target, no notebook runs above 0.66$\times$. No notebooks are sped up in the 10s target runtime as well. Unlike the 5s target runtime, 11 notebooks run faster than 0.11$\times$, but none run faster than 0.96$\times$.

In general, the slowdowns were significant. The maximum and geometric-mean slowdowns across the 4 target runtimes were: 0.0005$\times$ and 0.139, respectively. On the other hand, the maximum speedup was relatively small (2.10$\times$).

These conclusions are consistent with the experimental results of prior
work~\cite{dias}, but that prior work used much smaller input sizes and a more
narrow range (``approximately 2GB or less''). Here, we generalize that
conclusion across a much larger range of input sizes: \textasciitilde770 bytes
to \textasciitilde8 gigabytes.


We now investigate the one notebook that gets a significant speedup across any
target runtime: notebook 62. Figure~\ref{plt:per-cell-speedup-modin} presents a
cell-by-cell breakdown of absolute speedups of this notebook. It shows that the
first cell incurred a significant slowdown; a cell which, as is usual, primarily
consists of a call to \code{read\_csv}. This indicates that Modin's file loading
infrastructure worked slower than \code{pandas}. Nevertheless, Modin was able to
get an overall speedup, mainly by speeding up cell \#6 by almost 19s! This cell computes many string replacements like the following, with a
total of 13 calls to \code{str.replace()} or \code{.replace()}:

\begin{minted}[bgcolor=light-gray]{python}
vf2['NombreCliente'] = vf2['NombreCliente']
    .replace(['.*WAL MART.*','.*SAMS CLUB.*'],
    'Walmart', regex=True)
\end{minted}

\minihead{Dask} Now we turn our attention to Dask in
Figure~\ref{plt:speed-rel-dask}. Immediately we can observe that it
\textit{slows down all} notebooks across \textit{all} target runtimes. Secondly,
the ``speedups'' (i.e., the relative runtimes) are about the same across target
runtimes. 

\minihead{Koalas} The results for Koalas are shown in
Figure~\ref{plt:speed-rel-koalas}. Like Dask, Koalas \textit{slows down all}
notebooks across \textit{all} target runtimes. However, unlike Dask, Koalas
seems to scale better.

\minihead{Dias} We present the results for Dias in
Figure~\ref{plt:speed-rel-dias}. Similar to previous techniques, Dias does
\textit{not} speed up most of the notebooks. Also, contrary to the evaluation of
the paper~\cite{dias}, it slows down a significant portion of the notebooks on
the default target runtime (Figure~\ref{subfig:dias-default}). Nevertheless, we
see significant slowdowns \textit{only} on the default target runtime, which we
posit happens because of the small input size (something which agrees with
\cite{dias}). Furthermore, the slowdowns are much smaller than any of the other
techniques, with almost no notebook running below 0.9$\times$ on other
target runtimes. 

Moreover, Dias is the only technique to speed up notebooks on every target
runtime, and by a wide margin. The highest performance among the other
techniques comes from Modin, with 8 notebooks on the 20s target runtime. Dias
consistently speeds up more than 18 notebooks. Furthermore, the speedups seem to
scale (albeit slowly) as we increase the input size, which prior
work~\cite{dias} did not observe as they did not use different input sizes.

\minihead{Observations Across Pandas Optimization Techniques} First,
similar to \S\ref{sec:eval-cover}, our conclusions in this section are
consistent with findings of prior work~\cite{dias}. Again that prior evaluated
over a very limited set of snippets (i.e., only 3). Here we generalize these
observations for different input sizes and a larger set of notebooks.

Second, we make further observations across the alternatives. There is one
notebook on which all Pandas alternatives perform ``well'': notebook 92. Even
though it gets a slowdown across all techniques and all target runtimes (besides
Modin 20s), compared to \textit{other notebooks}, it is the one of fastest. One
peculiar observation is that Dias performs the relatively poorly on this
notebook.

We now show how \system{} can allow us to perform a fine-grained, cell-by-cell
comparison of the techniques by focusing on notebook 92. Both Dask and Koalas
massively speed up a cell that reads three CSV files, likely due to lazy
evaluation deferring the actual work of reading the CSV files to later cells.
For the next cell, which simply calls \code{DataFrame.info()}, Modin and Koalas
experience massive slowdowns, while Dask experiences a slight speedup, likely
because Dask's \code{.info()} is partial.\footnote{This means we don't get all
the information we get from \code{.info()} in \code{pandas}. We can't fix that
by evaluating it eagerly because it does not return anything. We would have to
evaluate (eagerly) the object we call it on (e.g.,
\code{evaluate\_eager(df).info()}). However, in Dask evaluating a dataframe
creates a \code{pandas.DataFrame} and not a Dask object. Given that we get
partial information, we could possibly consider it an \emph{Invalid Output}
issue (\S\ref{eval:reasons-fail}), but we considered this too strict since it does not cause a failure.} A
later cell applies a Python user-defined function (UDF) to a dataframe column,
and evaluates the result. Modin speeds this cell up, possibly by parallelizing
the call (prior work~\cite{dias} observed a similar speedup case because Modin
parallelized an \code{apply()} call). On the other hand, both Dask and Koalas
slow this cell down; Koalas only slightly, but Dask slows it down significantly.

\subsubsection{Memory \& Disk Usage}
\label{eval:mem-usg}


\begin{figure}[ht]
  \begin{subfigure}{0.95\columnwidth}
     \includegraphics[width=0.95\columnwidth]{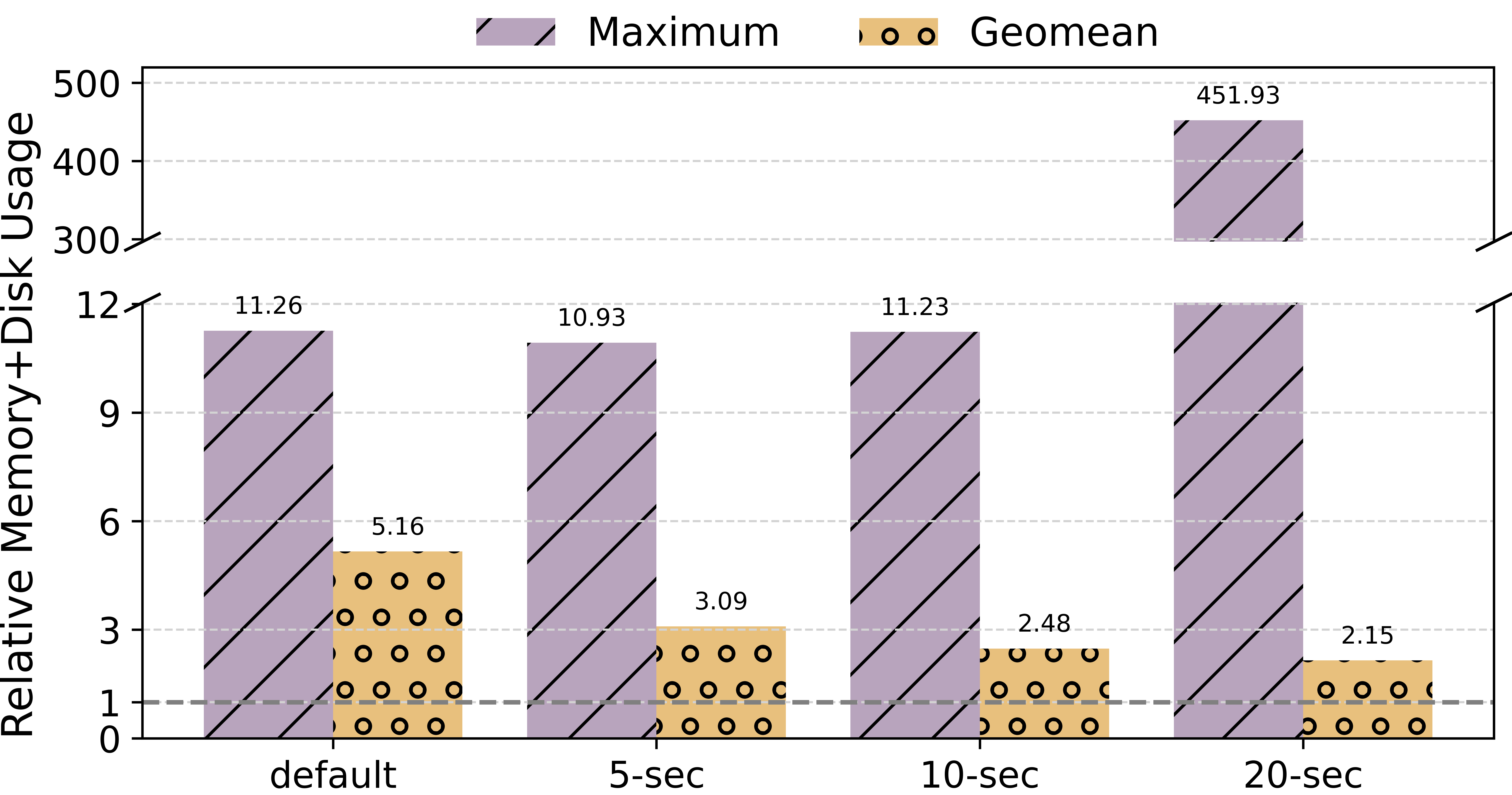}
    \caption{Modin memory consumption}
    \label{subfig:modin-mem}
  \end{subfigure}
  \begin{subfigure}{0.95\columnwidth}
    \includegraphics[width=0.95\columnwidth]{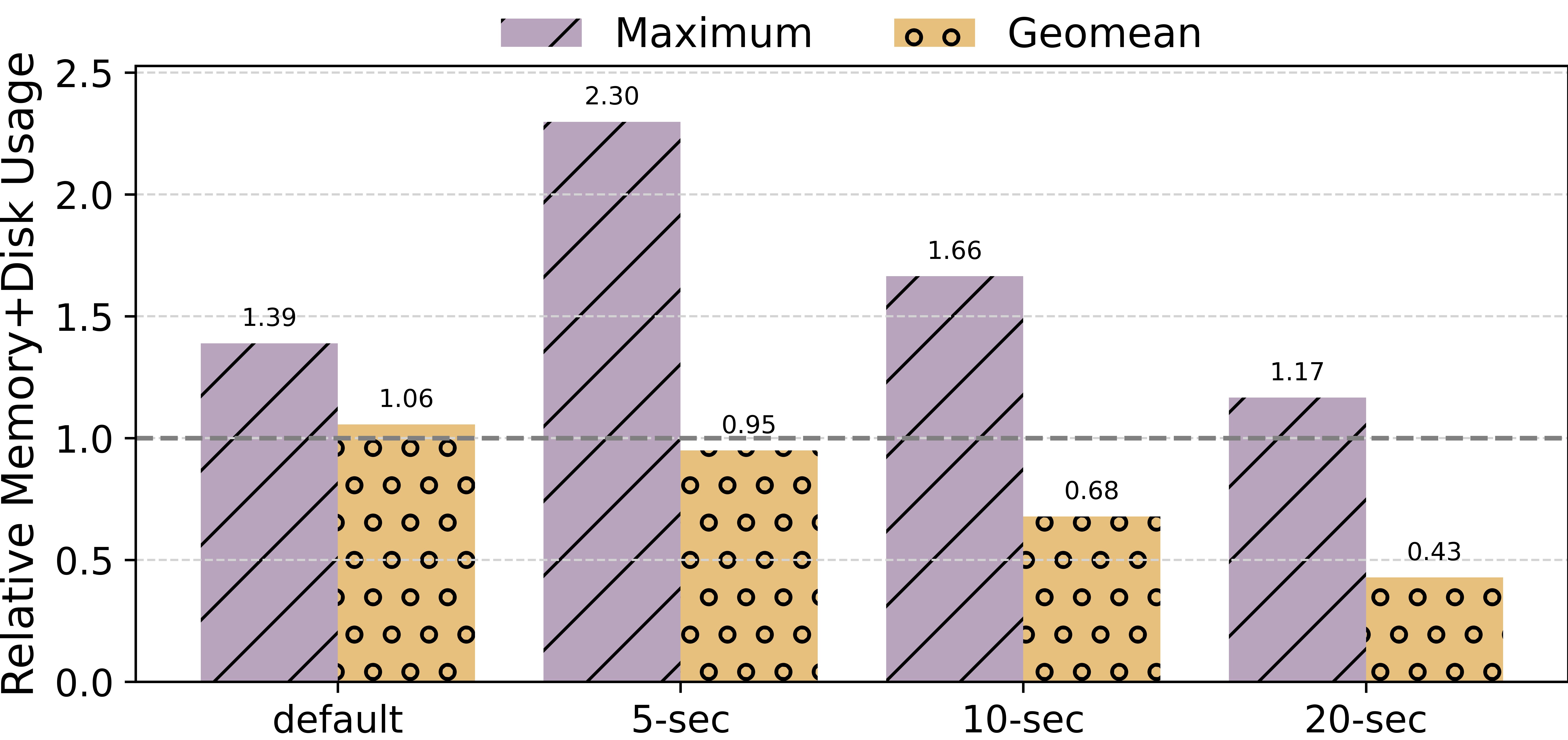}
    \caption{Dask memory consumption}
    \label{subfig:dask-mem}
  \end{subfigure}
  \begin{subfigure}{0.95\columnwidth}
     \includegraphics[width=0.95\columnwidth]{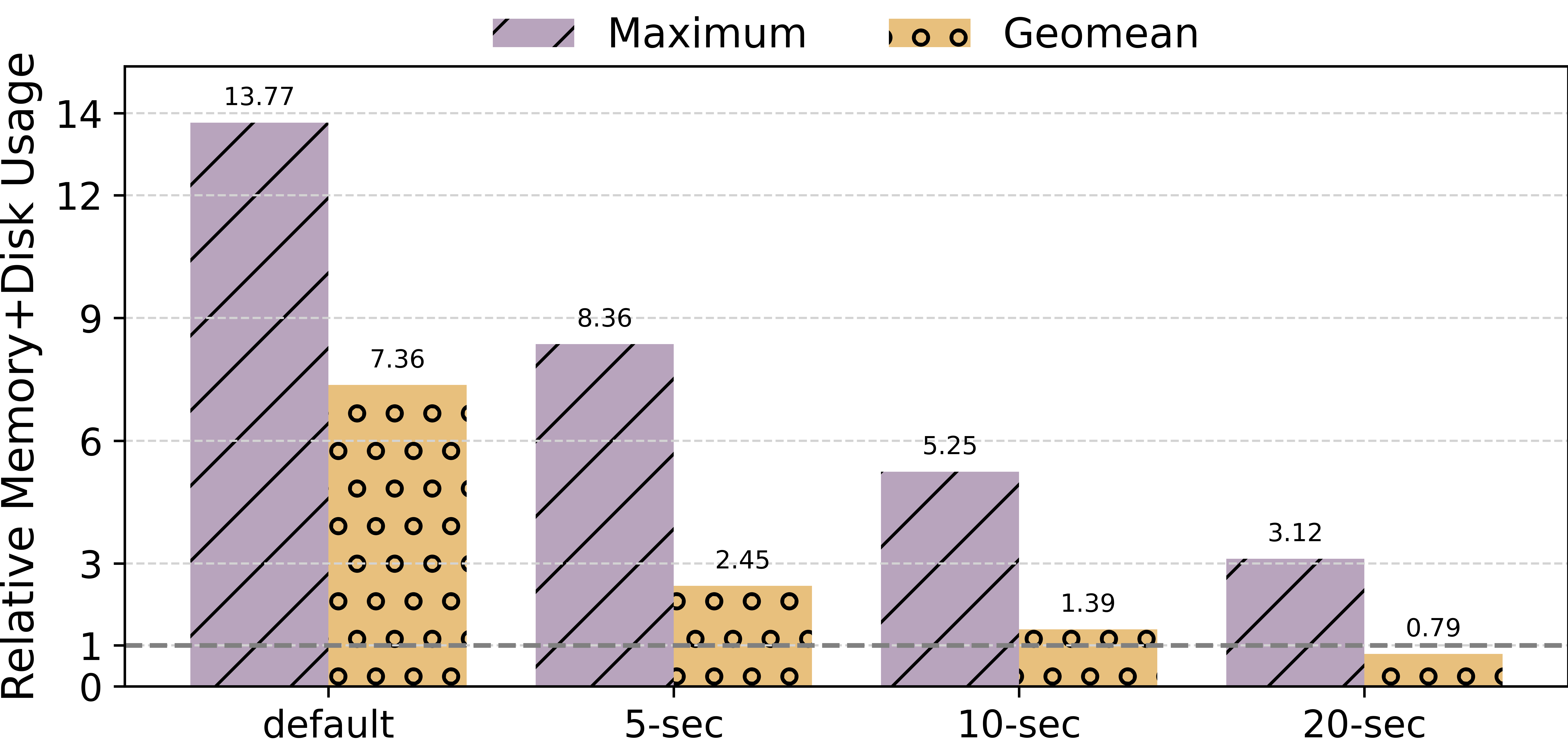}
    \caption{Koalas memory consumption}
    \label{subfig:koalas-mem}
  \end{subfigure}

  \caption{Maximum and geometric-mean \textit{relative} memory+disk usage of Modin, Dask, and Koalas compared to \code{pandas}. Dias
  is excluded because its usage is near-constant.}
  \label{plt:rel-mem}
\end{figure}

Figure ~\ref{plt:rel-mem} shows the \textit{relative} memory + disk usage of
runs of \system{} on Modin, Dask, Koalas, and Dias against runs of \system{} on
\code{pandas}. More specifically, to compute each geometric-mean $G$, for
a technique $T$ and target runtime $R$, we do the following:

\[
G_{T,R} = \text{GeoMean}\left(\left\{\frac{I_{T,R}}{I_{\code{pandas},R}} \, | \, I \in N_{T,R}\right\}\right)
\]

$I_{T,R}$ is the memory usage of running notebook $I$ with technique $T$ in the
runtime $R$. $N_{T,R}$ is the set of notebooks that ran, which depends $T$ and
$R$ (\S\ref{sec:eval-cover}). This computation shows how much more memory, on
average, each technique uses compared to \code{pandas}. We do a similar
computation for the maximum relative usage, by replacing GeoMean with Max which
computes the maximum value.

Here we see one of the most surprising results of the paper.
Figures~\ref{subfig:dask-mem} and ~\ref{subfig:koalas-mem} show that the memory
usage of both Dask and Koalas is \textit{inversely proportional} to the
input-data size! In fact, from the 5s target runtime and beyond, Dask uses, on
 average, \textit{less} memory than \code{pandas}. The same is true for Koalas on
 the 20s target runtime. While relative memory usage decreases, both
 maximum and geometric-mean memory usage for Dask and Koalas still increase as we go
 from the 5 to the 20 second target runtimes. The results highlight that Dask and
 Koalas \textit{scale} better than \code{pandas}, whereas it is less clear if Modin scales better or worse than
 \code{pandas}, with regards to memory. Modin's average memory usage relative to \code{pandas} decreases, but its worst case memory usage does not.

The worst case of memory usage of Modin is its biggest differentiator from other
Pandas optimization techniques. While (arithmetic) average memory usage is within reason
(\textasciitilde1.85GB on the 5s TR and \textasciitilde5.03GB on the 20s TR), peak memory usage of Modin changes from \textasciitilde3.63GB on
the 5s TR to \textasciitilde59.6GB on the 20s TR. On the same TRs over the same notebooks that Modin can run, \code{pandas} peaked at
\textasciitilde1.89GB and \textasciitilde7.81GB. Overall, Modin increased peak
memory usage by \textasciitilde$1.92\times$ and \textasciitilde$7.62\times$ for the
two respective TRs, and up to \textasciitilde$11.26\times$ and \textasciitilde$451.93\times$ for individual notebooks on the two respective TRs.
Unlike Dask and Koalas, Modin becomes worse
relative to \code{pandas} on larger TRs, with regards to peak memory usage. For Dask, on 5s and 20s
TRs the it yields a peak memory usage of
\textasciitilde1.11GB and \textasciitilde1.65GB while \code{pandas} yields
\textasciitilde1.35GB and \textasciitilde4.98GB on the same notebooks. Dask
transitioned from \textasciitilde$0.82\times$ to \textasciitilde$0.33\times$ the
peak memory usage of \code{pandas}. On the 20s TR, the Dask peak memory usage is lower than the \code{pandas} geometric-mean usage.
Koalas especially excels with scalability, having a
peak memory usage of \textasciitilde3.08GB and \textasciitilde3.11GB on the 5s
and 20s TRs.

Similarly to \S\ref{sec:eval-cover}, Dias predictably uses slightly more
memory than vanilla \code{pandas}, since it runs on top of it. The geometric
mean is almost the same, but the maximum memory consumption is
\textasciitilde$1.08\times$, which is almost the same across TRs because the techniques used by Dias are independent of the data size.

\section{Related Work}

\minihead{Pandas Collections} A concise presentation of prior work in Pandas
collections was presented in Table~\ref{tab:relwork_pandas}. Most existing
collections~\cite{distill_kaggle, datalore_dataset, data_science_looking_glass,
kgotrrent, analysis_millions_notebooks} are not
benchmarks (\S\ref{sec:back-defs}) as they are either not executable or not
available. The Dias notebooks~\cite{dias} form a benchmark \textit{and} use
real-world code, but they are too few (20) to be considered large and diverse. Modin's
evaluation~\cite{modin_downloads} does not use real-world code and is neither
executable nor large and diverse. In short, \system{} is the only benchmark that
fulfills all the requirements for a practical Pandas API benchmark.

\minihead{Relational Benchmarks} The standard set of relational benchmarks is
the TPC suite~\cite{tpc_summaries}. In Appendix~\ref{app:non-pandas} we have a
detailed argument of why TPC-H and other variants are not appropriate for a
Pandas API benchmark. In summary, they don't use the Pandas API, they don't
interact with the host language, and they don't include common dataframe use
cases (e.g., reading and writing from raw files). A related benchmark is
ProcBench~\cite{procbench}. As the name suggests, this is more of a procedural
than a relational benchmark, but it uses SQL variants that interoperate with
standard SQL, so we include it in this category. ProcBench still suffers from
the same problems because it uses the API of a conventional RDBMS (so, no
reading/writing of raw files, no interaction with a host language, no appearance
of the Pandas API).

\minihead{Numerical Benchmarks} We use the term ``numerical benchmark'' as a
generalization of any benchmark that deals explicitly with numeric data (be it
floating point or integer), including matrix and tensor benchmarks. Generally
these benchmarks are insufficient because the Pandas API makes heavy use of
other data types, and particularly strings (see \S\ref{sec:expl:input}). For the
same reason, \code{numpy} benchmarks~\cite{npbench,numpy_benches,
pythran_tests,numba_bench} are not sufficient, even though they are the closest
numerical benchmarks to the Pandas API.

\section{Conclusion}

We presented \system{}, the first benchmark for the Pandas API with real-world coverage. We described in detail how we collected the benchmark, as well as novel decisions we made to fit the special nature of the Pandas API and real-world code (e.g., our data scaling). We also explored the benchmark to show that it captures essential characteristics of the Pandas API and the dataframe model. Finally, we used the benchmark to compare Pandas optimization techniques, and identify surprising experimental observations.



\bibliographystyle{ACM-Reference-Format}
\bibliography{related}

\break

\appendix

\section{TPC Variants}
\label{app:non-pandas}


In this section we focus on TPC-H. Arguments on why other TPC variants are
unfitting follow similarly.

First, TPC-H does not test Pandas API coverage because it's written in SQL.
Second, TPC-H even if we try to translate it to Pandas API, it does not carry
over to dataframes as Pandas workloads do not translate well to SQL; Jindal et
al. \cite{magpie} found that about 60\% of Pandas method usage directly
corresponds to ``those methods that perform relational operations or operations
that are commonly supported by popular database systems,'' which leaves roughly
40\% unaccounted for.

Furthermore, TPC-H fails to capture common characteristics of a exploratory data
analysis (EDA) workflow, like: burst computations, reading and
writing (raw data) files, data cleaning, and inspection of
intermediate results. Reading and
writing is not covered as TPC-H only reads/writes data from/to the database, and
not from/to any files like CSV, Parquet, JSON, etc. Inspection is not tested
TPC-H issues whole, single queries. Finally, data cleaning, which is the removal
of data based on some characteristics identified after visualizing and
inspecting data, is not tested because that fundamentally requires an
interaction with the user, which TPC-H is not built to test. The closest to data
cleaning that we could find is the Old Sales Refresh Function (see Section~2.7
in the TPC-H specification \cite{tpch_spec}). However, this doesn't remove data
selectively to clean it, but rather ``removes rows from the ORDERS and LINEITEM
tables in the database to emulate the removal of stale or obsolete
information.''

Another important aspect of dataframe implementation (especially \code{pandas})
is their interoperability with the host programming language (e.g., Python),
which TPC-H simply does not test because it is all in SQL. Also, Pandas API code
frequently uses user-defined functions (UDFs), something TPC-H does not
benchmark because it has no user-defined functions.

To show this discrepancy between TPC-H workloads and Pandas API workflows in
practice, consider the case of one \code{pandas} alternative, Dask. When
compared with \code{pandas} on (the first) 7 queries from TPC-H, it speeds up
every single one of them, with a maximum speedup of $~4.6\times$
\cite{polars_bench_web,polars_bench_src}\footnote{We were able to replicate the
PolaRS benchmarks and we use the numbers from our experiments.}. However, when
compared with \code{pandas} on 3 commonly occurring EDA snippets, it failed to
run on one of them due to insufficient coverage, and slowed down the other 2,
with a maximum slowdown of $179\times$ \cite{dias}.

Other TPC variants are similarly unfitting because no variant includes any of:
coverage, burst computations, reading and writing (raw data) files, data
cleaning, inspection (or visualization) of intermediate results, interaction
with a host language. This is true for ProcBench
\cite{procbench} too, even though it includes procedural code and user-defined
functions.
\section{Complete Preparation Process}
\label{app:cleaning}

To fix, clean, and scale the notebooks, we adhered to the following algorithm (which was applied manually):

\begin{enumerate}
  \item \textbf{Setup:} Start in a Jupyter notebook with a \code{Python}
  environment with the target \code{pandas} version. Download the input data and change all file references in the notebook to refer to where the local input data lives.
  \item \textbf{Fixing Step:}
  Go through the notebook and minimally update code so the notebook can run. Section~\ref{sec:prep-fix} details different problems that may need to be fixed. Of particular importance: most notebooks try to get input files from a \code{/kaggle/} directory, which does not exist in typical Unix-like environments. These paths need to be updated to relative file system paths that actually exist.

  The details of the fixes to the problems from Section~\ref{sec:prep-fix} are detailed below.
  \begin{itemize}
    \item \textbf{Update Column Name:} Update the code to match the input data rather than updating the input data to match the code.
    \item \textbf{Replace With Equivalent Method:} Do a
    quick internet search for the nonexistent method. Most times we found online
    discussion concerning a replacement function call to use in newer versions
    of \code{pandas}, and update the notebook accordingly. Older versions of the
    \code{pandas} documentation itself can also provide equivalent replacements.
    For example, code that uses \code{.ix} is outdated, and old \code{pandas}
    documentation states that \code{.iloc} and \code{.loc} should be used instead.
    \item \textbf{Slice Object:} See if it is feasible to take a
    slice of the right hand side object with the same size as the dataframe column; if
    this is impossible remove the code by following the rules from the cleaning
    step.
  \end{itemize}

  Once any remaining errors are caused by libraries that will ultimately be
  excluded in the cleaning step, move on to the cleaning step.
  \item \textbf{Cleaning Step:} As opposed to applying fixes like in the previous step,
  to clean the notebooks, apply a set of rules. As a rule of thumb, if a rule produces code that
  assigns a named variable to itself (\code{df = df}) or assigns \code{\_} to a
  named variable (\code{\_ = df}), remove the assignment since these assignments
  are simply update object references in Python which is an inexpensive operation.

  Rules are justified if they preserve as much execution of \code{pandas} code
  as possible while removing all unrelated code, or expensive code that could
  sizably influence the runtime and/or memory usage of a notebook.

    \begin{itemize}
        \item Remove any \code{IPython} macros (prefixed with a \code{\%}).
        \item Remove any import that is unrelated to \code{pandas}, \code{numpy}, or
    the \code{Python} standard library.
        \item Remove any code that has nothing to do with \code{pandas} code, such
    as code that prints out the contents of directories on the file system.
        \item Remove any code that calls methods from a removed import. If the
    removed method takes \code{pandas} objects as arguments, we extract the
    \code{pandas} argument and assign it to \code{\_}.
        \item In the case that \code{pandas} calls into a plotting library (for
    instance, \code{df.plot.hist()}), keep the \code{pandas} code up to the
    point where a plotting method or attribute is accessed, and apply the above
    point to the method's arguments.
        \item If removed code assigns to a dataframe column, in general remove all
    following uses of that column by following the cleaning steps. An exception
    is if the original code assigns a transformation of a dataframe column to
    another (or even the same) column in a dataframe; in this case assign the
    original column to the new column and remove any following code that depends
    on the semantics of the new column according to the rules of the cleaning
    step.
        \item For loops, first see if enough code has been preserved that the loop
    header can run (possibly after modification). If the loop header can not be
    preserved, remove the loop. Otherwise, we apply the cleaning steps to the
    loop body, possibly replacing the body with a \code{pass} statement if the
    body becomes empty.
        \item For if statements, apply a similar approach to loops.
        \item For functions, if a parameter has a default value that depends on
    removed code, or a call site of the function passes in removed code as a
    corresponding argument, remove the parameter. Apply the cleaning steps to
    the function body; if the function body becomes empty remove the entire
    function by following the cleaning steps.
    \end{itemize}
  \item \textbf{Repeat} the fixing and cleaning steps until the notebook can run
  to completion after restarting the Jupyter environment.
  \item \textbf{Adaptation:}
    Replace the first \code{pandas} import with

    \code{exec(os.environ['PBENCH\_IMPORTS'])} (this may need to be proceeded by
    \code{import os}) and remove all subsequent \code{pandas} imports. Update
    all code that uses \code{pandas} to an equivalent version that would work if
    the only \code{pandas} import was \code{import pandas as pd}. For example, if the code originally had \code{import pandas.Series as Series} and then subsequent code like \code{Series()}, then we replace this subsequent code with \code{pd.Series()}
    
    Other changes may need to be made to ensure the execution of notebooks is
    suitable for a benchmark. These changes should be minimal and made to ensure
    consistent execution of a notebook. See Section~\ref{sec:prep-adapt} for a
    list of problems, with fixes detailed below.
    \begin{itemize}
        \item \textbf{Wrap With Evaluator:} Wrap with \code{evaluate\_eager} any code that would evaluate (or display) an intermediate result in the original notebook (i.e., before deleting irrelevant code).
        \item \textbf{Allow Repeated Execution:} Remove the smallest possible
        problematic segment of code by applying the cleaning step.
    \end{itemize}
  \item \textbf{Data Scaling:}
    Add in an additional cell at the beginning of the notebook (the first cell of any notebook that passed the data scaling step can be used as a template). This cell will be used to create a scaled version of the input data for this notebook. To scale the input, duplicate or remove rows until the number of rows in the dataframe has changed by the desired scale factor. Update the code template to refer to all
    input files used in the notebook and to account for other issues, such as
    file extensions or encodings. In the rest of the notebook, replace all
    references to input files in the notebook to the scaled version of the input
    (\code{"input.csv"} becomes \code{"input.scaled.csv"}).

    Run the notebook with scales of 0.3 and 2.1 and fix any issues that
    arise until the notebook runs correctly again. Common fixes
    are detailed below.
    \begin{itemize}
        \item \textbf{Update Scaled File:} Update how the scaled input file is generated in the new beginning cell; often this will concern deduplicating a column.
        \item \textbf{Replace With Runtime Value:} Update the code in the rest of the notebook to use runtime values instead of hardcoded values; often the runtime value will be the size of a list. In the case of downscaling bounds checks may need to be inserted.
    \end{itemize}

    To compute the scaling factor such that notebooks run for at least $X$ seconds, we run each notebook and calculate the ratio between the target time and the notebook runtimes. For notebooks with a ratio above 1, we increase their scale factor by slightly more than the calculated ratio. For ratios well below 1, we decrease their scale factor by the calculated ratio. We repeat until we converge on usable scale factors.
\end{enumerate}

\end{document}